\documentclass[a4paper,fleqn]{cas-dc}

\usepackage[comma]{natbib}
\renewcommand{\cite}{\citep}

\usepackage[utf8]{inputenc} 

\usepackage{graphicx}
\usepackage{lipsum}  
\usepackage{url}
\usepackage{fancyvrb}
\usepackage{xcolor}
\usepackage{hyperref}
\usepackage{silence}
\usepackage{multirow}
\usepackage{makecell}
\usepackage{adjustbox}
\usepackage{tabularx}
\usepackage{csvsimple}
\usepackage[inline]{enumitem}
\usepackage{amsfonts}

\usepackage{graphicx}
\usepackage{stfloats}   

\usepackage{epigraph}
\usepackage{listings}
\usepackage{amssymb} 
\usepackage{booktabs}
\usepackage{balance}

\usepackage{array}

\usepackage{caption}
\usepackage{subcaption}

\usepackage{cleveref}
\crefformat{section}{\S#2#1#3} 
\crefformat{subsection}{\S#2#1#3}
\crefformat{subsubsection}{\S#2#1#3}
\crefformat{appendix}{\S#2#1#3}

\usepackage[textsize=tiny]{todonotes}

\usepackage{xcolor}
\usepackage{fancyvrb}
\usepackage[most]{tcolorbox}

\tcbset{
  rqstyle/.style={
    colback=gray!15, colframe=black!50, fonttitle=\bfseries,
    coltitle=black, sharp corners
  },
  ansstyle/.style={
    colback=gray!5, colframe=black!30, fonttitle=\bfseries,
    coltitle=black, sharp corners
  }
}

\newtcolorbox{researchquestion}[2][]{rqstyle, title=Research Question~#2,#1}
\newtcolorbox{answer}[2][]{ansstyle, title=Answer to RQ#2,#1}

\usepackage[ruled,vlined,linesnumbered]{algorithm2e}
\usepackage{changepage}
\SetKw{KwBy}{by}
\SetKw{Push}{push}
\SetKw{Empty}{empty}
\SetKw{PushBack}{pushback}
\SetKw{Top}{top}
\SetKw{Pop}{pop}
\SetKwProg{Fn}{Function}{}{end}

\usepackage{pifont}
\newcommand{\cmark}{\ding{51}}%
\newcommand{\xmark}{\ding{55}}%

\newcommand{\journalcontent}[1]{}

\AtBeginDocument{%
  \providecommand\BibTeX{{%
    \normalfont B\kern-0.5em{\scshape i\kern-0.25em b}\kern-0.8em\TeX}}}

\newcommand{\ourtool}{{\sc Bullseye}}
\newcommand{\aflgo}{{\sc AFLGo}}
\newcommand{\windranger}{{\sc WindRanger}}
\newcommand{\firmafl}{{\sc Firm-AFL}}

\usepackage{amsthm}
\newtheorem{definition}{Definition}
\newtheorem*{definition*}{Definition}

\usepackage{enumitem}
\usepackage{tikz}
\usetikzlibrary{shapes.geometric,arrows.meta,positioning,calc,backgrounds,fit}

\makeatletter
\renewcommand{\fps@figure}{tp}
\makeatother

\begin{document}
\let\WriteBookmarks\relax
\def\floatpagepagefraction{1}
\def\textpagefraction{.001}

\shorttitle{BullsEye: Directed Firmware Fuzzing}    

\shortauthors{Ralli et al.}  

\title [mode = title]{BullsEye: Directed Firmware Fuzzing}  



%


\author[1, 2]{Lorenzo Ralli}[type=editor,
      auid=000,
      bioid=1,
      orcid=0009-0005-5472-894X
]
\ead{lorenzo.ralli@uniroma1.it}
\author[2]{Emilio Coppa}[type=editor,
      auid=000,
      bioid=2,
      orcid=0000-0002-8094-871X]
\ead{ecoppa@luiss.it}






\affiliation[1]{organization={Sapienza University of Rome},
            country={Italy}}
\affiliation[2]{organization={LUISS University},
            country={Italy}}











\begin{abstract}
The widespread adoption of Internet of Things (IoT) devices has expanded the digital attack surface, making firmware analysis critical for modern software security. A key security concern stems from the frequent reuse of third-party software components, a practice that often introduces known vulnerabilities into firmware images. Whether a given image actually exposes such a flaw is an open question, and public proof-of concept exploits make answering it urgent.

Directed Greybox Fuzzing (DGF), a technique that enables targeted exploration of specific binary locations, offers a promising solution for detecting such vulnerabilities. However, DGF has reached firmware only at function granularity, too coarse to aim at the vulnerable block itself.

This article presents \ourtool{}, the first DGF framework to schedule closed-source Linux-based firmware fuzzing by basic-block-level distance to user-specified targets. Our methodology combines static and dynamic analysis to enable DGF in the constrained firmware domain, focusing on vulnerabilities in reused third-party components. We introduce novel DGF heuristics that address limitations of traditional approaches. We compare \ourtool{} against four greybox-fuzzing baselines sharing its execution back-end, including reimplementations of \aflgo{} and \windranger{}, and against \textsc{Greenhouse}, a state-of-the-art firmware re-hosting framework. On 40 vulnerability sites across 32 firmware images, \ourtool{} reproduces every target within budget, against 35 for the strongest of the four baselines, and reduces Time-to-Exposure by a geometric mean of $9.5\times$ to $72.5\times$ over them; against \textsc{Greenhouse}, on the 18 targets its pipeline supports, \ourtool{} is faster by a geometric mean of $9.8\times$.
\end{abstract}




\hypersetup{pageanchor=false}
\maketitle
\hypersetup{pageanchor=true}


\section{Introduction}
\label{sec:intro}

The ever-increasing integration of embedded systems into everyday devices is expanding the digital attack surface at an unprecedented rate. Recent projections estimate that by 2034 the number of connected IoT devices will exceed 40~billion, up from 17~billion in 2024~\cite{arnott2025global}.
These systems span IP cameras, home routers, and critical industrial controllers, and are routinely deployed with a weak security posture dictated by cost and time-to-market pressures. The 2024 \textit{State of IoT Security} report recorded more than 900~million cyberattacks targeting IoT devices, a 114\% increase over the previous year~\cite{forescout2025threat}, driven in part by the absence of uniform security standards, which leads manufacturers to adopt ad hoc practices for each product domain~\cite{cirne2022iot}.

A dominant source of firmware vulnerabilities is the reuse of third-party components (TPCs), which implement security-critical operations such as HTTP parsing, authentication, cryptography, and file handling. Because TPCs are compiled once and shipped as part of the firmware binary, vendors rarely rebuild them against updated dependencies, leaving known vulnerabilities latent in deployed devices for years~\cite{chen2024autofirm}. When a public proof-of-concept exploit exists for a TPC vulnerability, the question is not whether the flaw is present but whether a specific firmware configuration exposes it. Answering this question efficiently requires focusing testing effort on the suspected vulnerability site rather than exploring the binary indiscriminately.

Coverage-guided greybox fuzzing~\cite{afl1,manes2019art} addresses the broader problem of automated vulnerability discovery by combining lightweight edge-coverage instrumentation with a mutation-driven feedback loop: test cases that exercise previously unseen control-flow edges are retained in a queue and mutated further, progressively exploring the program's input space without necessarily requiring source code or formal specifications. Directed greybox fuzzing (DGF)~\cite{dgf1} refines this paradigm for targeted analysis: instead of maximising coverage, the fuzzer assigns each test input an energy budget proportional to its estimated proximity to a set of user-specified target locations (typically the basic blocks associated with a known vulnerability), concentrating the campaign on execution paths likely to reproduce the flaw. DGF has demonstrated substantial reductions in time-to-exposure (TTE), the time to the first input triggering the vulnerability at the target, on source-available programs, motivating many successors~\cite{wang2024progress}.

Despite the evolution of DGF approaches, this methodology has not yet been extensively applied to firmware analysis. One barrier and three heuristic limitations stand in the way (detailed in Section~\ref{sec:motivation}). The barrier is source availability: existing DGF tools rely on compiler infrastructure (LLVM passes) to analyse the program and instrument it with distance information, so no directed campaign can start without a binary-level equivalent of that pipeline. The remaining three are not firmware-specific, but firmware's loop-structured code makes them acute. First, DGF concentrates effort on the conditional branches that guard access to the target, but identifying \emph{which} branches matter requires understanding control-flow structure; most existing approaches use a flow-insensitive analysis that fails silently on the loop-structured protocol parsers ubiquitous in firmware, missing the very conditions the fuzzer must solve. Second, measuring how far a test input is from the target by counting hops in the control-flow graph cannot distinguish an input whose input bytes are one character away from satisfying a hard comparison from one semantically far from it. Third, DGF shifts the campaign from broad exploration toward focused exploitation on a single global clock, so a critical branch reached late inherits an already-exploitative schedule with no budget to explore the newly discovered region, while the same campaign-global view pools energy on the targets easiest to reach and starves the rest.

We present \ourtool{}, a novel directed greybox fuzzer for closed-source IoT firmware. \ourtool{} is built on top of \firmafl{}~\cite{firmafl}, which internally uses QEMU for firmware
emulation, and addresses all four problems through a combination of binary-level static analysis, extended graph semantics, and a semantically richer distance metric, together with a set of mutation and bug-detection optimisations tailored to the emulation layer.

\paragraph{Contributions.}
In more detail, this article makes the following specific contributions:

\begin{itemize}

  \item \textbf{Binary-level DGF pipeline}. We design and implement a DGF pipeline that operates entirely on closed-source MIPS Linux-based firmware binaries. A static binary analysis pass extracts intra-procedural control-flow graphs, an inter-procedural call graph, and a call-site map without source code, and QEMU's dynamic translation layer is extended to accumulate per-input distance and branch-direction information at runtime and to refine the graphs as indirect call edges are observed.

  \item \textbf{Extended critical-branch analysis}. DGF must identify which conditional branches lie on the critical path to the target, i.e., branches whose outcome determines whether execution can ever reach the vulnerability. Existing approaches miss two branch classes common in firmware parsers: branches that guard loop iterations (where the loop must be re-entered to stay on the target path), and branches through which one successor must pass again to reach the target, while the other has a route around it. We refine the reachability condition with two further variants based on forward-only reachability, yielding a three-condition characterisation.

  \item \textbf{Satisfiability-adjusted distance}. We introduce a \emph{satisfiability} score that measures, for each critical branch, how close the current input is to satisfying that branch's comparison, derived from the operand values observed during execution. The score scales the structural distance, with per-input distance aggregated as a reciprocal-sum, so an input one byte-flip from solving a hard comparison ranks above one that is merely close in control-flow hops.

  \item \textbf{Per-branch annealing and target fairness}. We replace the single global cooling schedule shared by \aflgo{}~\cite{dgf1} and \windranger{} with an independent schedule per critical branch, starting when that branch is first reached. A branch discovered late thus receives a fresh exploration burst instead of inheriting an already-exploitative schedule. Inputs that bring the input closer to satisfying a hard comparison are also admitted to the corpus even without new coverage edges, and energy is weighted per target so that the targets easiest to reach do not starve the rest.

  \item \textbf{Optimisations.} \ourtool{} integrates several optimisations: (i)~a queue ordered by satisfiability-adjusted distance and per-target energy weight, so that each cycle begins from the most promising input; (ii)~grammar-aware mutation using the Grammar Mutator~\cite{grammar_mutators} and an LLM-assisted grammar extraction pipeline, with the bytes already matched to a comparison operand protected from destructive mutation; (iii)~a length-extension mutation that uses comparison-operand taint to clone copy-source bytes past the destination buffer bound; and (iv)~two overflow oracles, one tracking global-variable bounds from the binary's dynamic symbol table and one combining a heap chunk walk with allocation redzones and a copy-time bound check, both synthesising crash signals for silent corruptions that AFL's default oracle would miss.

  \item \textbf{Evaluation.} We evaluate \ourtool{} on 40 vulnerability sites in widely reused third-party components, across 32 real-world IoT firmware images, against four greybox-fuzzing baselines sharing its execution back-end, including reimplementations of \aflgo{} and \windranger{}, and against \textsc{Greenhouse}~\cite{jun2023greenhouse}, a state-of-the-art firmware re-hosting framework. \ourtool{} reproduces all 40 within budget, against 35 for the strongest of the four baselines, and is faster in time-to-exposure by a geometric mean of $9.5\times$ to $72.5\times$ over them; against \textsc{Greenhouse}, on the 18 targets its pipeline supports, \ourtool{} is faster by a geometric mean of $9.8\times$.

\end{itemize}

To encourage further research, we make\footnote{We plan to release the code after acceptance.} our contributions available at \url{https://github.com/BullsEyeDGF}.

\section{Background}
\label{sec:background}

This section introduces the technical concepts and prior frameworks that \ourtool{} builds upon. Section~\ref{sec:back:cgf} reviews coverage-guided greybox fuzzing, the CmpLog technique for comparison-guided mutation, the RedQueen colorization technique used in length-extension
mutation, and grammar-aware mutation techniques used for structurally constrained firmware inputs. Section~\ref{sec:back:dgf} presents directed greybox fuzzing, covering \aflgo{}'s seed-target distance formulation and \windranger{}'s deviation basic-block refinement.
Section~\ref{sec:back:firmware-fuzzing} surveys firmware fuzzing execution strategies and their limitations. Section~\ref{sec:back:firmafl} describes \firmafl{}, the execution engine on which \ourtool{} is built.

\subsection{Coverage-Guided Greybox Fuzzing}
\label{sec:back:cgf}

\begin{figure}[pos=htbp]
  \centering
  \includegraphics[width=0.8\linewidth]{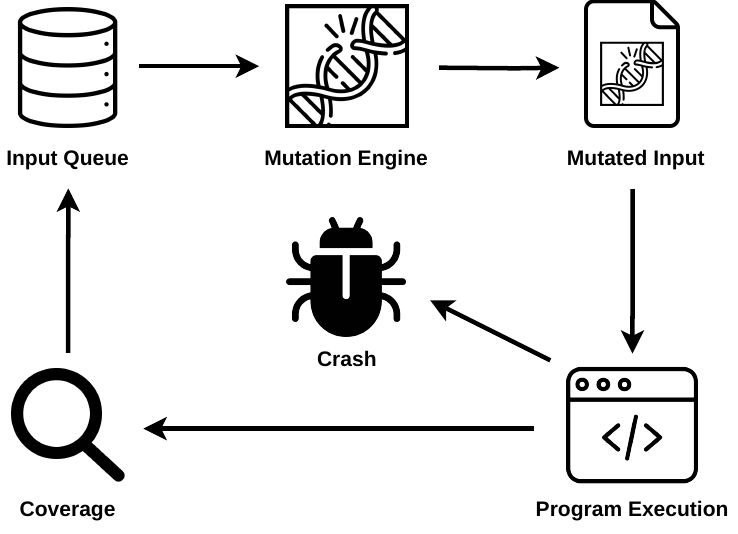}
  \caption{Coverage-Guided greybox fuzzing loop.}
  \label{fig:cgf}
\end{figure}

Coverage-guided greybox fuzzing (CGF) automatically generates test cases by combining lightweight instrumentation with a mutation-driven feedback loop~\cite{afl1,manes2019art}. American Fuzzy Lop (AFL), its reference implementation, maintains a queue of
seeds, the test inputs it selects for mutation and execution; seeds that trigger previously unseen control-flow edges are retained for further mutation (Figure~\ref{fig:cgf}).

\paragraph{Edge coverage bitmap.}
AFL instruments the program under test by injecting a small snippet before each basic block. At runtime, every transition from a predecessor block $p$ to a successor block $s$ updates a shared 64-KB \emph{coverage bitmap} at the index:
\[
  \texttt{idx} = \bigl((\texttt{pc}(p) \gg 1) \oplus \texttt{pc}(s)\bigr)
    \mathbin{\&} (\texttt{MAP\_SIZE}-1),
\]
where $\texttt{pc}(\cdot)$ denotes the program-counter value of the block. An execution is considered \emph{interesting} if any bitmap cell changes value relative to the currently known coverage. Because this hashing scheme maps many distinct edges to the same cell, it is an approximation; however, its $O(1)$ per-edge cost and compact representation make it practical for high-throughput fuzzing~\cite{afl1}.

\paragraph{Forkserver and power scheduling.}
To reduce process-initialisation cost, AFL uses a \emph{forkserver}: the program under test is started once, initialised to a stable state, and then forked for each test case. The number of mutations applied to each seed is controlled by a \emph{power scheduler} in the form of \emph{energy}, the
mutations it receives, assigning more to seeds that are faster to execute or that trigger unique coverage patterns. \emph{Culling} further prioritises the queue by maintaining a minimal favoured subset whose coverage collectively spans all known edges, allowing redundant seeds to be deprioritised and reducing wasted fuzzing effort.

\paragraph{Mutation stages.}
Deterministic mutation stages (bit and byte flips, arithmetic increments, known-interesting values, dictionary tokens) precede non-deterministic \emph{havoc} mutations that apply random combinations of byte operations~\cite{afl1}. The mutation pipeline concludes with \emph{trimming}, which removes unnecessary bytes, and \emph{splicing}, which combines portions of distinct seeds to generate new inputs.

\paragraph{QEMU-mode instrumentation.}
When source code is unavailable, AFL can instrument binaries at runtime via QEMU's Tiny Code Generator (TCG)~\cite{qemu}. During each Translation Block (TB) translation, AFL inserts a call to the \texttt{afl\_maybe\_log} helper, which records the current TB's contribution to the coverage bitmap.

\paragraph{Comparison-guided mutation: CmpLog.}
Many programs contain hard-coded string or numeric constants that guard deeper code paths. Random byte-level mutations have a negligible probability of satisfying a 4-byte magic-constant check by chance. CmpLog~\cite{cmplog}, based on the RedQueen technique~\cite{redqueen},
addresses this by instrumenting every comparison instruction and comparison function such as \texttt{strcmp} to log \emph{both operands} at runtime. When a logged operand appears in the current seed, the corresponding bytes are replaced with the complementary operand via a targeted Input-to-State (I2S) mutation. CmpLog is implemented in AFL++'s QEMU instrumentation layer~\cite{fioraldi2020afl++}, for x86 and ARM binaries only, and greatly improves penetration of input-validation barriers~\cite{redqueen}. \ourtool{} ports it to QEMU's MIPS front end.

\paragraph{Colorization and taint-based byte attribution.}
A complementary step in RedQueen~\cite{redqueen} is \emph{colorization}: before the I2S pass, the seed's bytes are replaced with random values while \emph{preserving its execution path}, one range at a time, splitting and retrying a range whenever coverage changes. The resulting variant is executed alongside the original with comparison logging enabled: an operand whose logged value differs between the two runs is input-derived, and the bytes carrying that value, located by scanning the seed, form its \emph{taint region}. Colorization thus yields byte-level attribution with no shadow-memory overhead. \ourtool{} reuses this taint region in the length-extension mutation (Section~\ref{sec:len-extend}) to grow the copied data past the destination buffer bound.

\paragraph{Grammar-Aware Mutation}
\label{sec:back:grammar}
Firmware input interfaces impose structural constraints on the data they accept: HTTP requests must contain a valid request line and well-formed headers, while environment-variable inputs follow a rigid \texttt{KEY=VALUE} format. Unrestricted byte-level mutation routinely violates these constraints, causing inputs to be rejected by the parser before reaching any vulnerability-relevant processing logic. Grammar-aware mutation restricts the fuzzer's output to structurally valid inputs, so executions survive the parser and reach deeper logic.

The Grammar Mutator~\cite{grammar_mutators} is an AFL++ plugin that accepts a user-defined context-free grammar and generates inputs by construction, guaranteeing syntactic validity. It
plugs into AFL's standard feedback loop: grammar-valid inputs that trigger new coverage enter the corpus for further grammar-constrained mutation. It maintains a parse tree for each queued input, and both generation and mutation operate on that tree; \ourtool{} backports AFL++'s custom-mutator interface to \firmafl{}'s AFL~2.52b base to host it.

\subsection{Directed Greybox Fuzzing}
\label{sec:back:dgf}

Coverage-guided greybox fuzzing maximises the fraction of the program explored, but provides no mechanism to concentrate effort on a specific code region. Directed greybox fuzzing (DGF) addresses this gap by reformulating the fuzzing objective: instead of maximising coverage, the fuzzer minimises a \emph{seed-target distance} metric that quantifies how close a given input's execution trace is to a set of user-specified target locations~\cite{dgf1}. DGF is particularly suited to scenarios in which the analyst already knows where a vulnerability may reside, such as crash reproduction, patch-regression testing, or verification of code paths involving third-party components with known CVEs. The remainder of this section describes the two DGF frameworks that most directly inform \ourtool{}'s design.

\subsubsection{\aflgo{}: Seed-Target Distance}
\label{sec:back:aflgo}

\aflgo{}~\cite{dgf1} is the first practical directed greybox fuzzer. It extends AFL with a compile-time static analysis phase and a distance-aware power scheduler to steer fuzzing toward a set of user-specified target locations.

\paragraph{Graph extraction.}
\aflgo{} instruments the program under test during compilation using an LLVM pass that constructs per-function
control-flow graphs (CFGs) and an inter-procedural call graph (CG). Basic blocks are identified by source-file-and-line-number coordinates. A Python postprocessor runs Dijkstra's shortest-path algorithm first over the CG, assigning each function a distance $d_f$ to the target functions, then over each function's CFG, assigning each block a distance $d_b$, with call-site blocks weighted by the $d_f$ of their callee; when several targets are given, per-target distances are combined by harmonic mean. 

\paragraph{Seed-target distance.}
Let $t$ be a target location and let $\mathcal{B}(q)$ denote the set of basic blocks executed by seed $q$.  \aflgo{} defines the \emph{basic-block-level distance} of a block $b$ to $t$ as:
\[
  d_b(t) = \begin{cases}
    0 & \text{if } b = t, \\
    d(b, t) & \text{otherwise,}
  \end{cases}
\]
where $d(b,t)$ is built from the \emph{call sites}, the blocks that invoke a function from which $t$
is reachable. Each call site $c$ contributes the CFG distance from $b$ to $c$, plus $c$'s own target
distance, defined as 10 times the function-level distance of its callee; $d(b,t)$ is the harmonic mean of these contributions.

\begin{figure}
    \centering
    \resizebox{0.7\columnwidth}{!}{%
    \begin{tikzpicture}[
      bb/.style={draw, rounded corners=2pt, fill=white,
                 font=\scriptsize, minimum width=1.5cm, minimum height=5mm, inner sep=2pt},
      tgt/.style={draw, rounded corners=2pt, fill=red!15,
                  font=\scriptsize, minimum width=1.5cm, minimum height=5mm, inner sep=2pt},
      off/.style={draw, rounded corners=2pt, fill=gray!18,
                  font=\scriptsize, minimum width=1.5cm, minimum height=5mm, inner sep=2pt},
      arr/.style={->, >=Stealth, semithick},
      sel/.style={->, >=Stealth, semithick, blue!65!black},
    ]
      \node[bb]  (entry) {entry,\ $d_b{=}4$};
      \node[bb,  below=3mm of entry]              (b1) {$b_1$,\ $d_b{=}3$};
      \node[bb,  below left=6mm and -5mm of b1]   (b2) {$b_2$,\ $d_b{=}2$};
      \node[off, below right=6mm and -5mm of b1]  (b3) {$b_3$,\ $d_b{=}\infty$};
      \node[bb,  below=3mm of b2]                 (b4) {$b_4$,\ $d_b{=}1$};
      \node[tgt, below=3mm of b4]                 (T)  {target,\ $d_b{=}0$};

      \draw[sel] (entry) -- (b1);
      \draw[sel] (b1) -- node[above left, font=\tiny]{T} (b2);
      \draw[arr, gray!55] (b1) -- node[above right, font=\tiny]{F} (b3);
      \draw[sel] (b2) -- (b4);
      \draw[sel] (b4) -- (T);

      \node[below=1mm of T, font=\small]
        {$D_{\textup{AFLGo}} = \dfrac{4+3+2+1+0}{5} = 2$};
    \end{tikzpicture}%
    }
    \caption{\aflgo{}'s seed-target distance. Each block is labelled with its shortest-path distance $d_b$ to the target (shaded red). For a seed following the highlighted trace $\langle \text{entry}, b_1, b_2, b_4, \text{target}\rangle$, $D_{\textup{AFLGo}} = 2$. Block $b_3$ (grey) has no path to the target; any seed executing it would receive $d_b = \infty$, inflating its score.}
    \label{fig:aflgo-distance}
\end{figure}
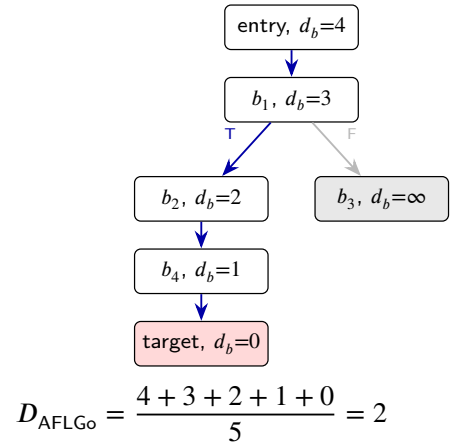

The seed-target distance is the arithmetic mean of $d_b(t)$ over $\mathcal{B}(q)$:
\[
  D_{\textup{AFLGo}}(q) = \frac{\displaystyle\sum_{b \in \mathcal{B}(q)} d_b(t)}{|\mathcal{B}(q)|}.
\]
A lower value of $D_{\textup{AFLGo}}(q)$ indicates that $q$'s execution trace passes through blocks that are, on average, closer to the target, as shown in Figure~\ref{fig:aflgo-distance}.

\paragraph{Simulated annealing power scheduler.}
\aflgo{} modulates AFL's power schedule with a \emph{simulated annealing} scheme. A global temperature $T_{\textup{g}}$ cools exponentially, $T_{\textup{g}} = 20^{-t/t_x}$, entering exploitation at $T_{\textup{g}} \le 0.05$, where $t_x$ is a user-specified time-to-exploitation budget. Writing $\tilde{D}(q) \in [0,1]$ for the seed distance min-max normalised against the current corpus, \aflgo{} sets
\[
    p(q) = \bigl(1 - \tilde{D}(q)\bigr)\,(1 - T_{\textup{g}}) + \tfrac{1}{2}\,T_{\textup{g}},
\]
and scales the energy AFL would have assigned by a factor $2^{10(p(q) - 1/2)}$. At $T_{\textup{g}} = 1$ every seed receives $p = 1/2$ and thus a factor of~1, leaving AFL's schedule untouched ({\emph{exploration phase}}); as $T_{\textup{g}} \to 0$ the factor spreads over $[1/32, 32]$ according to proximity ({\emph{exploitation phase}}). This scheduling strategy has been shown to outperform unguided fuzzing on crash-reproduction benchmarks~\cite{dgf1}.

\paragraph{Limitations.}
Three limitations of \aflgo{}'s formulation are central to this work.

First, the static analysis phase relies on LLVM IR, which is unavailable for closed-source binaries. Applying \aflgo{} to firmware therefore requires an alternative graph-extraction
mechanism~\cite{uaf}.

Second, the arithmetic mean aggregation treats every executed basic block equally. Blocks that are far from the target and have no path toward it nonetheless inflate the distance score of seeds that traverse them, introducing noise. \windranger{}~\cite{wind}, presented in Section~\ref{sec:back:windranger}, showed that restricting the metric to \emph{deviation basic blocks} substantially reduces this noise.

Third, the distance metric is purely structural. Two seeds that reach the same branch point are scored identically regardless of how closely their input bytes satisfy the branching condition at that point. This prevents the scheduler from distinguishing a seed that is one byte-flip away
from taking the correct branch from one that is semantically far from satisfying it.

\subsubsection{\windranger{}: Deviation Basic Blocks}
\label{sec:back:dbb}
\label{sec:back:windranger}

\windranger{}~\cite{wind} refines \aflgo{}'s distance metric by observing that most executed basic blocks are irrelevant to directional progress: they lie on straight-line code between branch points and cannot influence whether execution eventually reaches the target. Including them in the arithmetic mean of block distances dilutes the signal from the blocks that actually matter. \windranger{} addresses this by restricting the distance computation to a subset of blocks it calls \emph{Deviation Basic Blocks} (DBBs): branch points from which execution can either stay on a path toward the target or permanently deviate from it.

\paragraph{Formal definition.}
Let $G = (V, E)$ be the inter-procedural control-flow graph of the program, and let $\mathit{reach}(u, v)$ hold when there exists a path from $u$ to $v$ in $G$. \windranger{} defines
a block $b$ as a DBB with respect to target $t$ if and only if it satisfies two conditions:

\begin{enumerate}[label=(\arabic*)]
  \item $\mathit{reach}(b, t)$: the target is reachable from $b$; and
  \item $\exists s \in \mathit{succ}(b)$ such that $\neg\mathit{reach}(s, t)$: at least one
    successor of $b$ cannot reach $t$.
\end{enumerate}

\begin{figure*}[t!]
  \centering
  \begin{tikzpicture}[
    bb/.style={draw, rounded corners=2pt, fill=white,
               font=\scriptsize, minimum width=1.3cm, minimum height=5mm, inner sep=2pt},
    tgt/.style={draw, rounded corners=2pt, fill=red!15,
                font=\scriptsize, minimum width=1.3cm, minimum height=5mm, inner sep=2pt},
    dead/.style={draw, rounded corners=2pt, fill=gray!18,
                 font=\scriptsize, minimum width=1.3cm, minimum height=5mm, inner sep=2pt},
    dbb/.style={draw, rounded corners=2pt, fill=orange!25,
                font=\scriptsize, minimum width=1.3cm, minimum height=5mm, inner sep=2pt},
    arr/.style={->, >=Stealth, semithick},
    darr/.style={->, >=Stealth, semithick, dashed},
  ]
    \node[font=\small\bfseries, align=center] (la) at (0,0) {(a) if-else};
    \node[bb,  below=1mm of la]              (ae) {entry};
    \node[dbb, below=5mm of ae]              (ab) {$B$};
    \node[tgt, below left=7mm and 9mm of ab] (at) {target};
    \node[dead,below right=7mm and 9mm of ab](ad) {dead-end};

    \draw[arr] (ae) -- (ab);
    \draw[arr] (ab) -- node[above left, font=\tiny]{T} (at);
    \draw[arr] (ab) -- node[above right,font=\tiny]{F} (ad);
    \node[below=0mm of ad, font=\scriptsize, text=gray, align=center]
      {$\neg\mathit{reach}(\cdot,\,t)$};
    \node[left=2mm of ab, font=\scriptsize, text=green!50!black, align=center]
      {\cmark\ DBB};

    \begin{scope}[xshift=8.7cm]
      \node[font=\small\bfseries, align=center] (lb) at (0,0) {(b) loop-gating};
      \node[bb,  below=1mm of lb]              (be) {entry};
      \node[bb,  below=5mm of be]              (bb) {$B$};
      \node[tgt, below left=7mm and 9mm of bb] (bt) {target};
      \node[bb,  below right=7mm and 9mm of bb](bn) {loop head};

      \draw[arr]  (be) -- (bb);
      \draw[arr]  (bb) -- node[above left, font=\tiny]{T} (bt);
      \draw[arr]  (bb) -- node[above right,font=\tiny]{F} (bn);
      \draw[darr] (bn.east) to[out=0,in=0,looseness=1.6]
        node[right, font=\scriptsize, align=left]{~~back edge} (bb.east);
      \node[below=0mm of bn, font=\scriptsize, text=gray, align=center]
        {$\mathit{reach}(\cdot,\,t)$\\via back edge};
      \node[left=2mm of bb, font=\scriptsize, text=red!70!black, align=center]
        {\xmark\ not a DBB\\{\tiny(no deviating successor)}};
    \end{scope}
  \end{tikzpicture}
  \caption{\windranger{} deviation basic block classification. \textbf{(a)} In a standard if-else, block $B$ (orange) has a false-branch successor with no path to the target, so it is a DBB. \textbf{(b)} In a loop-gating pattern, both successors of $B$ can reach the target, as the false branch loops back through the same block on the next iteration, hence $B$ has no deviating successor and is not classified as a DBB. Section~\ref{sec:mot:dbb} elaborates on the consequences of this choice.}
  \label{fig:windranger-dbb}
\end{figure*}
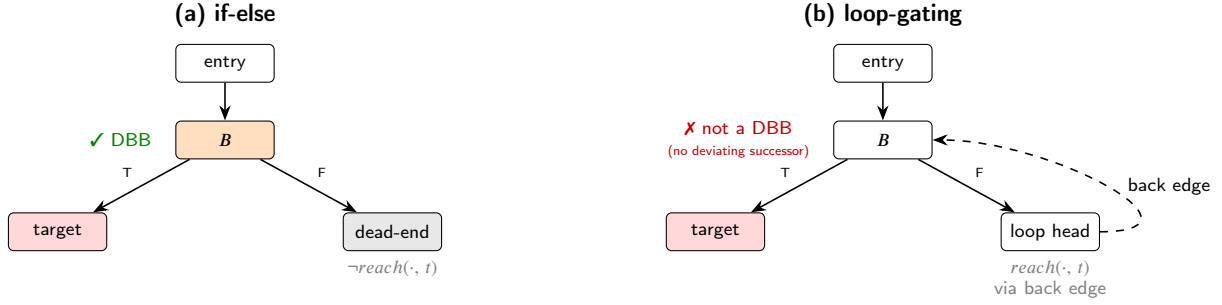

In other words, a DBB is a block that lies on some path to $t$ while having a successor that leaves all such paths; \windranger{} computes this statically using the SVF framework~\cite{svf} over LLVM IR. This yields the \emph{potential} DBBs of the program; for a seed, \windranger{} further retains only those blocks whose target-reaching successors are absent from its execution trace, so a block already traversed in the correct direction leaves the set. The resulting per-seed signal is binary: a block either still deviates, or drops out entirely. Figure~\ref{fig:windranger-dbb} contrasts a block that satisfies the conditions with one that does not.

\paragraph{Distance metric.}
Given the set $\mathcal{D}(q)$ of DBBs of a seed $q$, \windranger{} restricts the seed-target distance to DBBs. A seed's distance is the arithmetic mean of the distances of the DBBs in $\mathcal{D}(q)$, each weighted by a difficulty factor $\Psi$:
  \[
    D_{\textup{WR}}(q) = \frac{1}{|\mathcal{D}(q)|}
      \sum_{b \in \mathcal{D}(q)} d_b(t) \cdot \Psi(q, b)
  \]
where $\Psi(q,b)$ grows with the number of input bytes that influence the predicate of $b$. Seeds that do not hit any DBB receive a large default distance. Note that $\Psi$ scores the \emph{difficulty} of a predicate, not a seed's progress towards satisfying it: two seeds reaching $b$ with the same effective-byte count are indistinguishable, however close their operand values are to matching. \windranger{} thus inherits \aflgo{}'s third limitation, which \ourtool{} addresses in Section~\ref{sec:sat-score}.

\paragraph{Seed management and scheduling.}
To help seeds hit DBBs, \windranger{} computes a \emph{cb\_mask}: a bitmask over seed positions identifying bytes whose mutation is likely to affect a DBB condition, obtained by probing each position and observing which branch variables change. Masked positions are mutated with higher probability, concentrating effort where DBB predicates can be influenced. Seeds are prioritised per DBB: the closest seed covering each DBB is promoted to a favoured queue. Both behaviours are stage-dependent: \windranger{} alternates between an exploration and an exploitation stage, switching once every DBB has been hit often enough relative to the least-hit block. The trigger is therefore hit frequency, and energy is still governed by \aflgo{}'s single global annealing clock.

\paragraph{Limitations in the firmware context.}
\windranger{}'s conditions for DBBs capture the prototypical divergence point but miss two classes of branch that are common in firmware binaries and that \ourtool{} identifies as distinct DBB classes. The motivation and formal treatment of these extended classes are given in
Sections~\ref{sec:mot:dbb} and~\ref{sec:dbb-classifier} respectively. Furthermore, \windranger{}'s static analysis relies on LLVM IR and therefore cannot be applied to closed-source firmware without modification.

\subsection{Firmware Fuzzing}
\label{sec:back:firmware-fuzzing}

Applying coverage-guided greybox fuzzing to firmware requires solving a fundamental execution problem: the binary to test is compiled for an embedded processor (commonly MIPS or ARM), depends on device-specific peripherals, and cannot run directly on the analyst's host machine. Two broad
strategies have been explored for Linux-based images~\cite{wright2021challenges}. \emph{Full-system emulation}
approaches (e.g., TriforceAFL~\cite{triforceafl}, Firmadyne~\cite{firmadyne}) boot the complete firmware image inside QEMU, preserving environmental fidelity at the cost of low throughput (a few executions per second) due to software-based memory management and device emulation overhead. \emph{Hybrid user-mode} approaches (e.g., \firmafl{}~\cite{firmafl},
\textsc{EquAFL}~\cite{equafl}) migrate the process under test to a faster user-mode emulator once initialised, recovering AFL-comparable throughput while retaining the environment needed to boot real firmware images.

Despite this progress, all existing frameworks treat the firmware binary as an opaque execution artifact: the fuzzer receives an edge-coverage bitmap and a crash signal, but no information about how close an input is to a specific vulnerability site, so the directed dimension is still largely missing from the firmware fuzzing literature. \ourtool{} addresses this by building a directed greybox fuzzing layer on top of \firmafl{}'s execution engine, using the binary-level analysis that the absence of source (Section~\ref{sec:back:aflgo}) forces.

\subsection{\firmafl{}: Augmented Process Emulation}
\label{sec:back:firmafl}

\firmafl{}~\cite{firmafl} is one of the first high-throughput greybox fuzzers for IoT firmware. It targets Linux-based firmware images for embedded processors, whose services ship as stripped ELF binaries without source, extracted from device flash images and executed inside an emulated system.

\paragraph{Augmented Process Emulation.}
Prior firmware fuzzers relied on full-system QEMU emulation for all execution, which incurs high
overhead from software-based memory management and device emulation. \firmafl{} addresses this through \emph{Augmented Process Emulation}: the firmware is booted inside Firmadyne~\cite{firmadyne},
a full-system QEMU instance with a custom instrumented kernel. Once the binary under test is fully initialised, DECAF's Virtual Machine Introspection (VMI)~\cite{virt3} detects the fork point, and subsequent execution of the binary migrates to a user-mode QEMU instance. System calls and page faults are handled by selectively re-entering full-system mode. This hybrid approach achieves throughput comparable to AFL's user-mode fuzzing while preserving the environmental fidelity needed to boot real firmware images~\cite{firmafl}.

\paragraph{Input channels.}
\firmafl{} delivers inputs to the firmware binary through two channels that reflect its two primary interfaces:
\begin{itemize}
  \item \textit{HTTP socket channel}: the fuzzer intercepts the \texttt{read} system call on the listening socket and replaces the received data with the current test case. This channel is used for firmware binaries that serve an HTTP-based management interface.
  \item \textit{CGI environment-variable channel}: many firmware web servers invoke the application logic as a CGI process. In this model, the HTTP server passes request parameters through POSIX environment variables~\cite{posix_env} such as \texttt{REQUEST\_METHOD}, \texttt{QUERY\_STRING}, and \texttt{HTTP\_COOKIE}. \firmafl{} intercepts the process-environment setup and injects fuzz-controlled key-value pairs before execution begins.
\end{itemize}

\paragraph{Coverage and instrumentation.}
Coverage is collected via AFL's QEMU-mode edge bitmap in the user-mode QEMU instance. Branch transitions within the binary under test are recorded by the standard \texttt{afl\_maybe\_log} helper, described in Section~\ref{sec:back:cgf}. Because the binary is not recompiled, all static analysis must be performed at the binary level.

\section{Motivation}
\label{sec:motivation}


Firmware images deployed in IoT devices increasingly rely on third-party components to implement security-critical functionality such as HTTP parsing, authentication, and cryptographic operations. Because these components are compiled once and shipped as closed-source binaries,
manufacturers rarely rebuild them against updated dependencies, leaving known vulnerabilities latent within deployed devices for extended periods~\cite{wu2024your,chen2024autofirm}. When a public proof-of-concept exploit exists for such a vulnerability, the question is not
whether the flaw is present but whether a specific firmware configuration exposes it. Answering this question with general-purpose coverage-guided greybox fuzzing is inefficient:
it explores the binary indiscriminately, spending the majority of its budget on execution paths far from the vulnerability site.

Directed greybox fuzzing (Section~\ref{sec:back:dgf}) addresses this inefficiency. Rather than treating all edges as equally worth exploring, a DGF tool assigns each seed an energy
budget proportional to its estimated proximity to a set of designated target locations, typically the basic blocks associated with a known vulnerability. Seeds closer to the target receive more mutation effort, concentrating the campaign on execution paths likely to reproduce or trigger the flaw. \aflgo{}~\cite{dgf1} demonstrated substantial reductions in time-to-exposure (TTE) on source-available programs, motivating follow-on work such as \windranger{}~\cite{wind}, Beacon~\cite{beacon}, and
SelectFuzz~\cite{luo2023selectfuzz}.

Despite this progress, DGF has not been applied extensively to firmware analysis. The obstacles are of two different kinds. One is a genuine \emph{barrier to entry} that has kept basic-block-level DGF out of firmware altogether (\ref{p1}); the other three are \emph{limitations of DGF's guidance heuristics} that are not firmware-specific, but that firmware makes acute (\ref{p2}--\ref{p4}).

\begin{enumerate}[label=\textbf{P\arabic*}, ref=P\arabic*]
  \item \label{p1}
    \textbf{Closed-source binary analysis.}
    The DGF pipeline requires control-flow and call graphs together with compile-time-instrumented binaries, but IoT firmware is overwhelmingly distributed without source.

  \item \label{p2}
    \textbf{DBB classifier incompleteness.}
    The deviation basic block (DBB) classification proposed by \windranger{} is blind to the loop-structured code that dominates firmware protocol parsers, systematically missing the branches that gate the sink, the unsafe operation the target marks.

  \item \label{p3}
    \textbf{Structural-only distance.}
    Existing DGF distance metrics count control-flow hops but carry no information about how close a concrete input is to satisfying the branch predicates along the way.

  \item \label{p4}
    \textbf{Campaign-global energy allocation.}
    A single temperature spanning the entire run penalises seeds that first reach a late-stage guard deep in the campaign, when exploitation pressure is already maximal and the local input space has not yet been explored. The same campaign-global view extends across targets: a single distance signal pools energy on the targets that are easiest to reach and starves the harder
    ones, a fairness problem that follows from aggregating every target into one score.
\end{enumerate}

The remainder of this section elaborates each problem with concrete examples drawn from representative firmware.

\subsection{Closed-source Binary Analysis}
\label{sec:mot:binary}
DGF tools overwhelmingly assume source availability: \aflgo{}~\cite{dgf1}, \windranger{}~\cite{wind}, Beacon~\cite{beacon}, and SelectFuzz~\cite{luo2023selectfuzz} each instrument the program under analysis at compile time to emit a coverage bitmap and to embed the precomputed distance values that steer energy scheduling. UAFuzz~\cite{uaf} is a notable exception: it performs binary-level directed fuzzing via QEMU instrumentation and IDA-assisted graph extraction, but it targets Use-After-Free vulnerabilities in general x86 binaries and is not designed for the embedded Linux firmware setting we address. IoT firmware violates the source-availability assumption in practice. Vendors ship stripped ARM or MIPS binaries, often packed or encrypted, with no accompanying build system. Binary rewriting can recover a control-flow graph, but it cannot reliably insert the distance-accumulation hooks that DGF requires, and disassembly errors in position-independent or hand-optimised code propagate into incorrect distance values that silently mislead the fuzzer. A firmware-capable DGF pipeline must therefore perform graph extraction, target identification, and instrumentation entirely on the binary, and remain robust to the imprecision inherent in static binary analysis rather than propagating it into the distance metric.

\subsection{DBB Classifier Incompleteness}
\label{sec:mot:dbb}

Restricting distance computation and runtime tracking to deviation basic blocks (DBBs), as \windranger{} does (Section~\ref{sec:back:dbb}), concentrates the distance metric on branches that decide whether the target is reached. Under such a scheme, the quality of the DBB classifier sets a ceiling on how accurately the directed fuzzer can measure progress.

\windranger{}'s classifier tests only set membership: a block $b$ is a DBB if $b$ lies in the target's backward-reachable set and at least one successor $s$ does not (Figure~\ref{fig:windranger-dbb}). This captures the prototypical fork-in-the-road pattern but cannot distinguish \emph{reaches the target via a direct forward path} from \emph{reaches the target by looping back through this branch again}. The condition therefore produces false negatives on three loop-structured patterns common in firmware protocol parsers. DAFL~\cite{dafl} independently corroborates this limitation, observing that \emph{DBBs can be a bad representative when they are within a loop}; its own design forgoes DBBs entirely, replacing CFG-based distances with semantic relevance scores derived from source-level program slicing over a Def-Use Graph. This solution does not transfer to closed-source firmware binaries, where the inter-procedural data-flow analysis that program slicing requires loses precision without compiler IR.

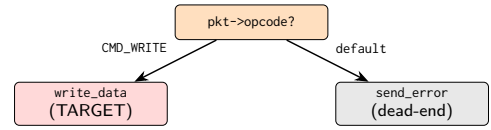
\begin{figure}
\centering
\resizebox{0.75\columnwidth}{!}{%
\begin{tikzpicture}[
  bb/.style={draw, rounded corners=2pt, fill=white, font=\scriptsize,
             minimum width=2.2cm, minimum height=5mm, inner sep=2pt, align=center},
  tgt/.style={draw, rounded corners=2pt, fill=red!15, font=\scriptsize,
              minimum width=2.2cm, minimum height=5mm, inner sep=2pt, align=center},
  exit/.style={draw, rounded corners=2pt, fill=gray!18, font=\scriptsize,
               minimum width=2.2cm, minimum height=5mm, inner sep=2pt, align=center},
  dbb/.style={draw, rounded corners=2pt, fill=orange!25, font=\scriptsize,
              minimum width=2.2cm, minimum height=5mm, inner sep=2pt, align=center},
  arr/.style={->, >=Stealth, semithick},
]
  \node[dbb] (sw) {\texttt{pkt->opcode?}};
  \node[tgt, below left=6mm and 1mm of sw]  (tg) {\texttt{write\_data}\\(TARGET)};
  \node[exit,below right=6mm and 1mm of sw] (er) {\texttt{send\_error}\\(dead-end)};

  \draw[arr] (sw) -- node[above left,  font=\scriptsize]{\texttt{CMD\_WRITE}} (tg);
  \draw[arr] (sw) -- node[above right, font=\scriptsize]{\texttt{default}}    (er);
\end{tikzpicture}%
}
\caption{Control-flow graph of the listing from the baseline pattern. The switch (orange) has a \texttt{default} arm with no path to TARGET, so the DBB condition is satisfied and \windranger{} identifies the critical branch.}
\label{fig:mot:switch}
\end{figure}

\paragraph{Baseline: Switch on opcode, one arm dead-ends.}
\windranger{}'s DBB condition is designed for branches where at least one successor cannot reach the target. A firmware command dispatcher is representative:

\begin{lstlisting}[language=C, label={lst:switch}]
switch (pkt->opcode) {
  case CMD_WRITE:
    write_data(pkt->buf, pkt->len); /* TARGET */
    break;
  default:
    send_error(conn, ERR_UNKNOWN_CMD);
    break;
}
\end{lstlisting}

The \texttt{CMD\_WRITE} branch reaches the vulnerable \texttt{write\_data} sink; the \texttt{default} branch sends an error response and returns without touching the target. Because the default successor has no path to the target, the DBB condition is satisfied and the switch is correctly classified as a DBB by both \windranger{} and \ourtool{}
(Figure~\ref{fig:mot:switch}).

The three patterns below all share the same symptom: \windranger{} does not mark a critical branch as a DBB. This follows from the strict definition of DBB, which does not account for the loop structures common in real-world firmware images.

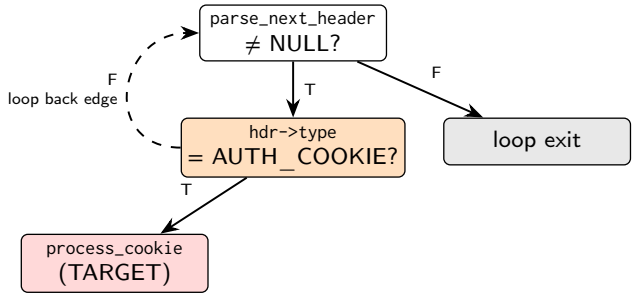
\begin{figure}
\centering
\resizebox{\columnwidth}{!}{%
\begin{tikzpicture}[
  bb/.style={draw, rounded corners=2pt, fill=white, font=\scriptsize,
             minimum width=2.0cm, minimum height=5mm, inner sep=2pt, align=center},
  tgt/.style={draw, rounded corners=2pt, fill=red!15, font=\scriptsize,
              minimum width=2.0cm, minimum height=5mm, inner sep=2pt, align=center},
  exit/.style={draw, rounded corners=2pt, fill=gray!18, font=\scriptsize,
               minimum width=2.0cm, minimum height=5mm, inner sep=2pt, align=center},
  dbb/.style={draw, rounded corners=2pt, fill=orange!25, font=\scriptsize,
              minimum width=2.0cm, minimum height=5mm, inner sep=2pt, align=center},
  arr/.style={->, >=Stealth, semithick},
  darr/.style={->, >=Stealth, semithick, dashed},
  tainted/.style={draw=teal!60, dashed, rounded corners=4pt, inner sep=3pt},
]
  \node[bb]  (lh)  {\texttt{parse\_next\_header}\\$\neq$ NULL?};
  \node[dbb, below=6mm of lh]  (lb)  {\texttt{hdr->type}\\$=$ AUTH\_COOKIE?};
  \node[tgt, below left=6mm and -3mm of lb]  (tg)  {\texttt{process\_cookie}\\(TARGET)};
  \node[exit,below right=6mm and 6mm of lh] (ex)  {loop exit};

  \draw[arr] (lh) -- node[right, font=\tiny]{T} (lb);
  \draw[arr] (lh) -- node[above right, font=\tiny]{F} (ex);
  \draw[arr] (lb) -- node[above left, font=\tiny]{T} (tg);
  \draw[darr](lb.west) to[out=180, in=180, looseness=1.8]
    node[left, font=\tiny, align=right]{F\\loop back edge} (lh.west);


\end{tikzpicture}%
}
\caption{Control-flow graph of the listing from Pattern~1. The inner conditional (\texttt{hdr->type == AUTH\_COOKIE}, orange) has both successors able to reach the target: the true branch reaches the target directly, the false branch loops back and can reach it on the next iteration. \windranger{}'s condition therefore does not classify this block as a DBB.}
\label{fig:mot:loop-gating}
\end{figure}

\paragraph{Pattern~1: Protocol-header enumeration loop.}
Consider the following excerpt, representative of HTTP servers found in D-Link and Trendnet firmware images:

\begin{lstlisting}[language=C, label={lst:loop-gating}]
while ((hdr = parse_next_header(buf, &pos))) {
    if (hdr->type == AUTH_COOKIE) {
        process_cookie(hdr->value); /* TARGET */
        break;
    }
}
\end{lstlisting}

The inner conditional (\texttt{hdr->type == AUTH\_COOKIE}) has two successors: the true branch reaches the target directly, and the false branch loops back to the outer while condition, which can reach the target on the next iteration. Because both successors can reach the target, \windranger{}'s DBB condition is not satisfied and the block is not classified as a DBB
(Figure~\ref{fig:mot:loop-gating}). Yet, this branch is the only gate on reaching the target: its operand is precisely what the fuzzer must learn to manipulate.

\medskip

The key issue is that the branch controls the iteration at which the target becomes reachable, rather than whether the target is reachable at all. \windranger{}'s DBB definition does not capture this iterative gating behavior.

\paragraph{Pattern~2: Bounded-retry authentication gate.} 
Another common pattern arises in the login services of router and IP camera web interfaces, which permit a bounded number of authentication retries. The privileged handler is dispatched at the \emph{top} of the retry loop, guarded by a flag that the previous iteration's credential check sets:

\begin{lstlisting}[language=C, label={lst:retry}]
int attempts = 0, authenticated = 0;
do {
  if (authenticated)
    launch_admin_handler(req); /* TARGET */
  char *c = cred;
  do { c++; } while (*c && *c != ':');
  authenticated = verify_credentials(cred);
} while (++attempts < MAX_RETRIES &&
         read_next_request(req));
\end{lstlisting}

\begin{figure}
\centering
\resizebox{\columnwidth}{!}{%
\begin{tikzpicture}[
  bb/.style={draw, rounded corners=2pt, fill=white, font=\scriptsize,
             minimum width=2.2cm, minimum height=5mm, inner sep=2pt, align=center},
  tgt/.style={draw, rounded corners=2pt, fill=red!15, font=\scriptsize,
              minimum width=2.2cm, minimum height=5mm, inner sep=2pt, align=center},
  exit/.style={draw, rounded corners=2pt, fill=gray!18, font=\scriptsize,
               minimum width=2.2cm, minimum height=5mm, inner sep=2pt, align=center},
  dbb/.style={draw, rounded corners=2pt, fill=orange!25, font=\scriptsize,
              minimum width=2.2cm, minimum height=5mm, inner sep=2pt, align=center},
  arr/.style={->, >=Stealth, semithick},
  darr/.style={->, >=Stealth, semithick, dashed},
]
  \node[bb]  (au)  {\texttt{authenticated?}};
  \node[tgt, above left=5mm and -4mm of au]  (tg)  {\texttt{launch\_admin\_handler}\\(TARGET)};
  \node[bb,  below=6mm of au]  (sb)  {\texttt{c++}};
  \node[dbb, below=5mm of sb]  (sl)  {\texttt{*c \&\& *c != ':'}\,?};
  \node[bb,  below=5mm of sl]  (ve)  {\texttt{verify\_credentials(cred)}};
  \node[bb,  below=5mm of ve]  (dw)  {\texttt{++attempts < MAX}\\$\wedge$\,\texttt{read\_next\_request?}};
  \node[exit, below right=6mm and -6mm of dw]  (ex) {loop exit};

  \draw[arr] (au) -- node[above, font=\tiny]{T} (tg);
  \draw[arr] (au) -- node[right, font=\tiny]{F} (sb);
  \draw[arr] (tg) to[out=-90, in=160] (sb.north west);
  \draw[arr] (sb) -- (sl);
  \draw[darr](sl.west) to[out=180, in=180, looseness=2.4]
    node[left, font=\tiny, align=right]{T\\inner\\back edge} (sb.west);
  \draw[arr] (sl) -- node[right, font=\tiny]{F} (ve);
  \draw[arr] (ve) -- (dw);
  \draw[darr](dw.east) to[out=0, in=0, looseness=1.15]
    node[right, font=\tiny, align=left]{T\\retry\\back edge} (au.east);
  \draw[arr] (dw) -- node[below right, font=\tiny]{F} (ex);
\end{tikzpicture}%
}
\caption{Control-flow graph of the listing from Pattern~2. The target is dispatched at the loop top; the inner credential-scan latch (\texttt{*c != ':'}, orange) sits below it. Both its successors reach the target only by traversing the outer retry back edge, never forward. \windranger{}'s condition therefore does not classify this block as a DBB.}
\label{fig:mot:retry}
\end{figure}
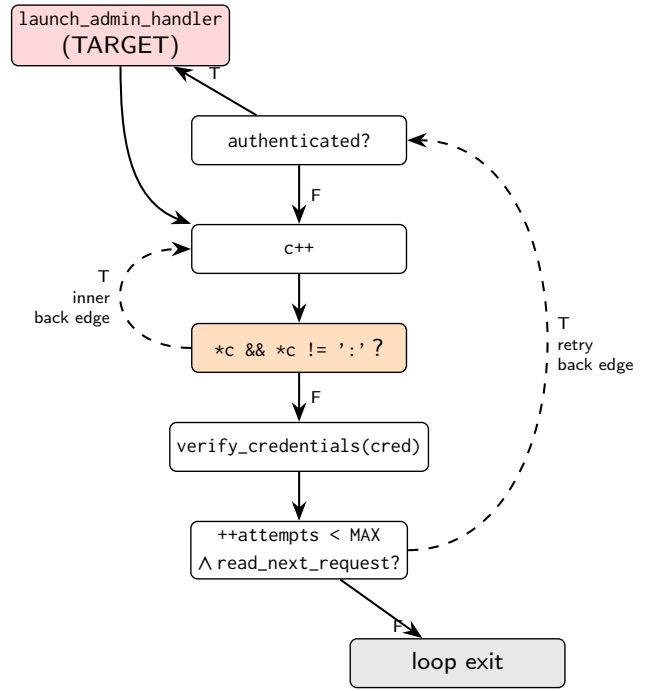

The inner scan latch (\texttt{*c \&\& *c != ':'}), the block holding the inner loop's back edge, has two successors: the continue-scan branch loops back to read more of the token, and the done branch falls through to \texttt{verify\_credentials}. The handler, however, is dispatched at the top of the loop, reached only once a credential verifies and control loops back, so both successors reach it only by traversing the outer retry back edge. \windranger{}'s condition is therefore not satisfied and the latch is not classified as a DBB (Figure~\ref{fig:mot:retry}). Yet, this latch gates every route to the handler: reaching it means going back around the loop, and its operand is precisely what the fuzzer must learn to satisfy.

\begin{figure}
\centering
\resizebox{\columnwidth}{!}{%
\begin{tikzpicture}[
  bb/.style={draw, rounded corners=2pt, fill=white, font=\scriptsize,
             minimum width=2.2cm, minimum height=5mm, inner sep=2pt, align=center},
  tgt/.style={draw, rounded corners=2pt, fill=red!15, font=\scriptsize,
              minimum width=2.2cm, minimum height=5mm, inner sep=2pt, align=center},
  exit/.style={draw, rounded corners=2pt, fill=gray!18, font=\scriptsize,
               minimum width=2.2cm, minimum height=5mm, inner sep=2pt, align=center},
  dbb/.style={draw, rounded corners=2pt, fill=orange!25, font=\scriptsize,
              minimum width=2.2cm, minimum height=5mm, inner sep=2pt, align=center},
  arr/.style={->, >=Stealth, semithick},
  darr/.style={->, >=Stealth, semithick, dashed},
]
  \node[bb]  (b)   {\texttt{len++}};
  \node[dbb, below=6mm of b]   (l)   {\texttt{buf[len] != '\textbackslash 0'}\,?};
  \node[bb,  below=6mm of l]   (crc) {\texttt{crc16(buf,len) == expected}\,?};
  \node[tgt, below left=6mm and -8mm of crc]  (tg)  {\texttt{apply\_config}\\(TARGET)};
  \node[exit, below right=6mm and -8mm of crc] (ex) {fail / exit};

  \draw[arr] (b) -- (l);
  \draw[darr](l.west) to[out=180, in=180, looseness=2.2]
    node[left, font=\tiny, align=right]{T\\scan\\back edge} (b.west);
  \draw[arr] (l)   -- node[right, font=\tiny]{F} (crc);
  \draw[arr] (crc) -- node[above left, font=\tiny]{T} (tg);
  \draw[arr] (crc) -- node[above right, font=\tiny]{F} (ex);
\end{tikzpicture}%
}
\caption{Control-flow graph of the listing from Pattern~3. The sink lies \emph{after} the scan loop; the scan latch (\texttt{buf[len] != '\textbackslash 0'}, orange) must terminate to reach it. Both its successors reach the target: the exit branch via the checksum test, the back-edge after one more iteration, so \windranger{}'s condition does not classify this block as a DBB.}
\label{fig:mot:scan}
\end{figure}
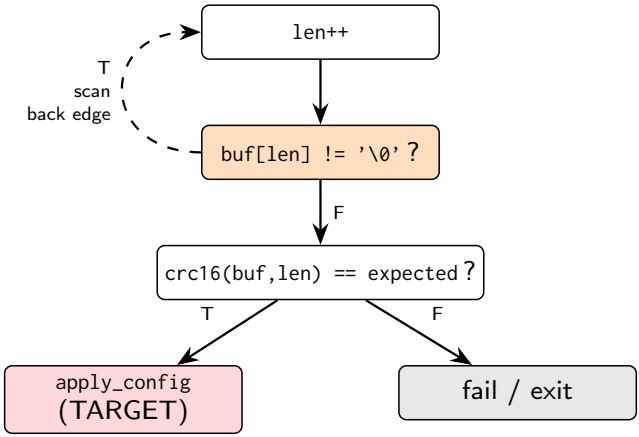

\paragraph{Pattern~3: Checksum-gated configuration copy.}
Configuration handlers in router web interfaces routinely measure a field, verify its integrity, and only then copy it, placing the vulnerable copy \emph{after} the measuring loop. The following is representative of the NVRAM settings-write paths in SOHO router firmware:

\begin{lstlisting}[language=C, label={lst:scan}]
size_t len = 0;
do { len++; } while (buf[len] != '\0');
if (crc16(buf, len) == expected)
  apply_config(buf, len);  /* TARGET */
\end{lstlisting}

The scan latch (\texttt{buf[len] != '\textbackslash 0'}) has two successors: the continue-scan branch reads another byte, and the done branch leaves the loop for the checksum test that guards \texttt{apply\_config}. Both reach the target: the done branch through the checksum test, the continue branch after another scan iteration. \windranger{}'s DBB condition is therefore not satisfied and the block is not classified as a DBB (Figure~\ref{fig:mot:scan}). Yet this latch is the input-termination gate on the whole sink: the target stays unreachable until the field is null-terminated within bounds, and producing it is exactly what the fuzzer must learn.

\begin{table}[t]
\centering
\caption{DBB classification for representative firmware code patterns (WR = \windranger{}, BE = \ourtool{}; brackets give the firing condition).}
\label{tab:dbb-patterns}
\small
\begin{tabular}{lcc}
\toprule
Code pattern & WR & BE \\
\midrule
Explicit if-else, one branch dead-ends     & \cmark\ {\scriptsize[2a]} & \cmark\ {\scriptsize[2a]} \\
Switch on opcode, one arm dead-ends   & \cmark\ {\scriptsize[2a]} & \cmark\ {\scriptsize[2a]} \\
Protocol-header enumeration loop           & \xmark                    & \cmark\ {\scriptsize[2c]} \\
Bounded-retry authentication gate          & \xmark                    & \cmark\ {\scriptsize[2b]} \\
Checksum-gated configuration copy          & \xmark                    & \cmark\ {\scriptsize[2c]} \\
\bottomrule
\end{tabular}
\end{table}

\paragraph{Summary.} 
Table~\ref{tab:dbb-patterns} summarises all patterns and their classification under \windranger{}'s single-condition definition (2a) versus the three-condition definition (2a, 2b, and 2c) used in this work (Section~\ref{sec:dbb-classifier}).

\subsection{Structural-Only Distance}
\label{sec:mot:distance}

Even with a correct set of DBBs, the distance metric determines how effectively those blocks guide the campaign. \aflgo{} computes the arithmetic mean of shortest-path distances across all executed blocks. \windranger{} restricts this average to DBBs and drops a DBB from a seed's set once that seed navigates it correctly. Both distances nonetheless remain structural: \windranger{}'s $\Psi$ term scales a DBB's contribution by how many input bytes affect its predicate, but neither metric reflects how close a given seed's operand values are to satisfying it. DAFL~\cite{dafl} takes a step beyond structural distance by replacing the control-flow graph with a Def-Use Graph (DUG) derived from inter-procedural data-flow analysis: seed distances are computed over data-dependency paths rather than control-flow hops, yielding a more semantically meaningful distance metric. The DUG is built by an LLVM pass over source-level intermediate representation, however, so it inherits the applicability limit noted in Section~\ref{sec:mot:dbb}.

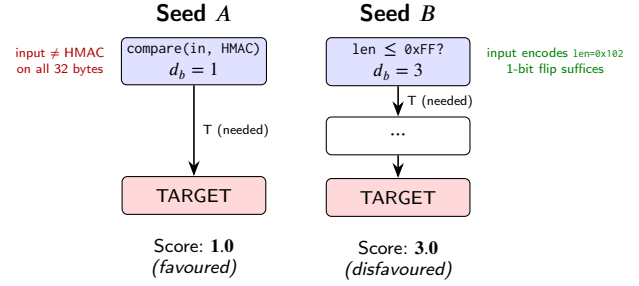
\begin{figure}
\centering
\begin{tikzpicture}[
  bb/.style={draw, rounded corners=2pt, fill=white, font=\scriptsize,
             minimum width=1.9cm, minimum height=5mm, inner sep=2pt, align=center},
  tgt/.style={draw, rounded corners=2pt, fill=red!15, font=\scriptsize,
              minimum width=1.9cm, minimum height=5mm, inner sep=2pt, align=center},
  cur/.style={draw, rounded corners=2pt, fill=blue!12, font=\scriptsize,
              minimum width=1.9cm, minimum height=5mm, inner sep=2pt, align=center},
  arr/.style={->, >=Stealth, semithick},
]
  \node[font=\small\bfseries, align=center] (la) at (0,0) {Seed $A$};
  \node[cur, below=1mm of la]  (aB) {\texttt{compare(in, HMAC)}\\$d_b = 1$};
  \node[tgt, below=12mm of aB]  (aT) {TARGET};
  \draw[arr] (aB) -- node[right, font=\tiny]{T (needed)} (aT);
  \node[left=1mm of aB, font=\tiny, text=red!70!black, align=center]
    {input $\neq$ HMAC\\on all 32 bytes}; 
  \node[below=2mm of aT, font=\scriptsize, align=center, text=black]
    {Score: $\mathbf{1.0}$\\\textit{(favoured)}};

  \begin{scope}[xshift=2.7cm]
    \node[font=\small\bfseries, align=center] (lb) at (0,0) {Seed $B$};
    \node[cur, below=1mm of lb]  (bB) {\texttt{len $\leq$ 0xFF?}\\$d_b = 3$};
    \node[bb, below=4mm of bB]   (b2) {$\cdots$};
    \node[tgt, below=3mm of b2]  (bT) {TARGET};
    \draw[arr] (bB) -- node[right, font=\tiny]{T (needed)} (b2);
    \draw[arr] (b2) -- (bT);
    \node[right=1mm of bB, font=\tiny, text=green!50!black, align=center]
      {input encodes \texttt{len=0x102}\\1-bit flip suffices}; 
    \node[below=2mm of bT, font=\scriptsize, align=center, text=black]
      {Score: $\mathbf{3.0}$\\\textit{(disfavoured)}};
  \end{scope}
\end{tikzpicture}

\caption{Seed~$A$ is one hop from the target but must satisfy a 32-byte HMAC comparison, which is virtually impossible by random mutation. Seed~$B$ is three hops away but requires only a 1-bit mutation. \aflgo{} and \windranger{} assign seed~$A$ higher energy than seed~$B$.} 
\label{fig:mot:semantic}
\end{figure}

To see why relying solely on structural distance is a problem in the firmware context, consider two seeds stalled at different DBBs on the way to a vulnerable \texttt{memcpy} in a firmware
CGI handler, as shown in Figure~\ref{fig:mot:semantic}. Both take the wrong branch:

\begin{itemize}
  \item \textbf{Seed~$A$} is one structural hop from the target. The guard compares the input against a 32-byte HMAC token derived from a per-session secret; no byte of seed~$A$ matches any byte of the required token.

  \item \textbf{Seed~$B$} is three structural hops away. The guard checks \texttt{content\_length $\leq$ 0xFF}; seed~$B$'s input encodes \texttt{content\_length = 0x102}, which differs from the satisfying range by a single bit.
\end{itemize}

\aflgo{} and \windranger{} assign seed~$A$ higher energy because it is structurally closer. In practice, seed~$A$ has a negligible probability of satisfying its guard through random mutation, whereas seed~$B$ requires a single bit-flip. Figure~\ref{fig:mot:semantic} visualises this disparity: structural distance can be a poor proxy for expected mutation cost.

What is missing is a measure of satisfiability proximity: a continuous assessment of how close the current input's runtime comparison operands are to the values that would satisfy the branch predicate. \windranger{}'s $\Psi$ weight scores the predicate, not the seed: a seed matching 31 of the 32 required bytes and one matching none receive identical treatment. This coarseness prevents the fuzzer from recognising incremental progress and from directing extra mutation effort toward seeds on the verge of satisfying a hard comparison.

The absence of a satisfiability gradient also limits the splice operator: AFL selects splice partners uniformly at random, whereas a principled strategy would pair a seed that fails at a given DBB with the corpus seed that most closely satisfies that DBB's predicate. Without a continuous satisfiability score, this pairing cannot be computed.

\subsection{Campaign-Global Energy Allocation}
\label{sec:mot:annealing}

\aflgo{} schedules fuzzing energy using simulated annealing: a global temperature $T(t)$ that decreases monotonically from~1 (uniform exploration) towards~0 (concentrated exploitation), crossing the exploitation threshold $0.05$ at a configurable time $t_x$. \windranger{} and later DGF tools retain this design. Hence, all DBBs share the same temperature regardless of when they were first individually reached.

For targets protected by a sequential chain of guards, this global schedule creates a temporal inequality. Suppose a target requires satisfying two DBBs, $d_1$ and $d_2$, in order. In a 45-minute campaign with $t_x$ set to its full length, where the fuzzer first reaches $d_2$ at minute~40, the remaining exploration budget for $d_2$ is five minutes, compared to the 40 minutes over which $d_1$ was explored. Yet $d_2$ has just been discovered: the fuzzer has had no opportunity to characterise the local input space around it. Subjecting it immediately to maximum exploitation pressure, when the temperature is near zero, is counterproductive.

The inequality is particularly harmful for multi-stage authentication logic or deeply nested protocol state machines in firmware, where satisfying late-stage DBBs depends on having first solved earlier ones. The natural response to discovering a new DBB is a fresh burst of exploration energy: a per-DBB temperature that starts high and cools independently of the global campaign clock.

A second, spatial inequality appears when a campaign specifies several targets at once. \aflgo{} supports multiple sinks by aggregating their per-target distances into a single score, but aggregation is silent on fairness. A single global distance signal pools energy on whichever targets are easiest to reach: seeds that arrive at a nearby, frequently hit sink crowd out those making slow progress toward a harder one, and the harder targets starve, receiving the least energy exactly when they most need directed exploration. Neither \aflgo{}'s aggregation nor the schedule \windranger{} inherits from it corrects this; energy follows the path of least resistance rather than the target most in need.

\subsection{Discussion}
\label{sec:mot:summary}

Of the four problems analysed in this section, only the first is a barrier to entry. Because IoT vendors ship stripped binaries without source, the compile-time instrumentation that current source-level DGF tools depend on is unavailable, and no directed campaign can even start (\ref{p1}). The remaining three are of a different kind: they are not barriers but limitations of DGF's guidance heuristics, present in source-available programs as well, that firmware's iterative, loop-structured code makes acute. The DBB classifier that determines \emph{which} branches the fuzzer treats as meaningful decision points is blind to the loop-structured idioms that dominate firmware protocol parsers, silently omitting the very branches that guard vulnerable sinks (\ref{p2}).  The distance metric compounds this: counting control-flow hops tells the fuzzer how many branches separate a seed from the target, but says nothing about how close a given seed is to satisfying them (\ref{p3}). And the campaign-global energy allocation means that a DBB discovered near the end of the campaign is immediately subjected to maximum exploitation pressure, with no time to explore the local input space around it, while in multi-target runs the same global accounting lets easily reached sinks monopolise energy and starves the harder ones (\ref{p4}). In short: \ref{p1} explains why DGF has not reached firmware at basic-block granularity, while \ref{p2}--\ref{p4} explain why porting an existing DGF tool onto a firmware image, even after \ref{p1} is cleared, would still produce poor results.

\section{\textbf{\textsc{Bullseye}}}
\label{sec:approach}

\begin{figure*}[pos=htbp]
  \centering
  \includegraphics[width=1.0\linewidth]{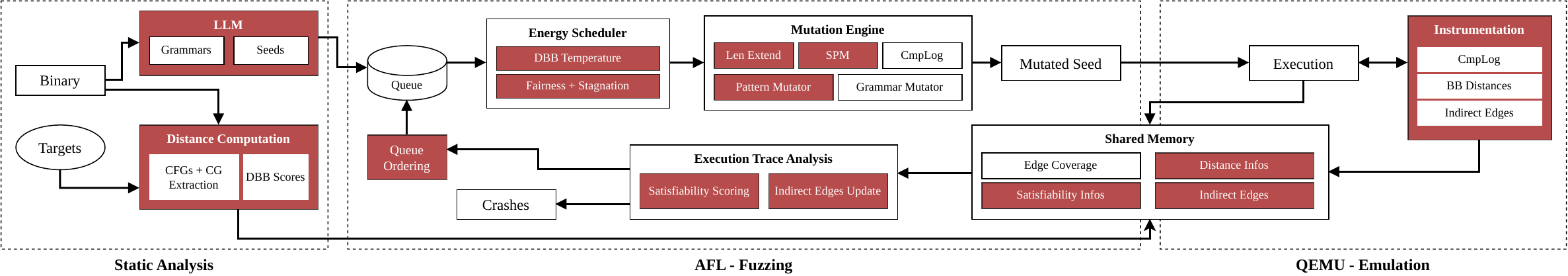}
  \caption{Architectural overview of \ourtool{}.}
  \label{fig:overview}
\end{figure*}

\ourtool{} is a directed greybox fuzzing framework for firmware analysis that addresses the four problems identified in Section~\ref{sec:motivation}.  Figure~\ref{fig:overview} gives an
architectural overview. The framework operates in two phases. In the \emph{static analysis phase}, a disassembler script extracts basic-block ranges and call-site information from the closed-source binary; a graph construction pass synthesises intra-procedural control-flow graphs and an inter-procedural call graph; and the distance computation pass classifies DBBs, computes per-block distances, and serialises the result to a file loaded by the fuzzer at startup. In the \emph{dynamic fuzzing phase}, QEMU-based instrumentation hooks record distances and branch-direction bits in the shared-memory region consulted by the scheduler; the mutator applies pattern-aware and comparison-guided mutations (Section~\ref{sec:mut-opt}); and the scheduler combines structural
distance with runtime satisfiability scores to allocate energy.

\ourtool{} tackles each problem through a dedicated contribution:

\begin{itemize}
  \item \textbf{Binary-level DGF pipeline} (\ref{p1}, Section~\ref{sec:binary-pipeline}): extract control-flow and call graphs from closed-source binaries, instrument firmware emulation to track distance at runtime, and refine the graphs dynamically as indirect edges are observed during execution.

  \item \textbf{Extended DBB classification} (\ref{p2}, Section~\ref{sec:dbb-classifier}): identify loop-gating and mandatory re-entry branches that \windranger{}'s classification ignores, introducing two new conditions.

  \item \textbf{Satisfiability-adjusted distance} (\ref{p3}, Section~\ref{sec:sat-score}): compute a continuous satisfiability measure from runtime comparison operands and fold it into the distance metric, allowing \ourtool{} to rank seeds that nearly satisfy a predicate above structurally closer seeds that do not.

  \item \textbf{Power scheduling and multi-target fairness} (\ref{p4}, Section~\ref{sec:scheduling}): give each newly discovered DBB its own independent exploration budget with a private cooling clock, decoupled from the global campaign temperature, and balance energy across targets so that easily reached ones do not starve the rest, reviving culled seeds when coverage stalls.
\end{itemize}

The remainder of this section presents each contribution in more detail. The optimisations listed in Section~\ref{sec:intro}, namely distance-aware queue ordering, grammar-aware and length-extension mutation, and the two overflow oracles, are bound to the emulation layer that realises them and are therefore presented in Section~\ref{sec:implementation}.

\subsection{Binary-Level DGF Pipeline}
\label{sec:binary-pipeline}

Addressing \ref{p1} requires replacing compile-time instrumentation with a pipeline that operates entirely on the closed-source binary. \ourtool{} structures this around three conceptual stages: static graph extraction, initial distance scores computation, and runtime graph refinement.

\paragraph{Static graph extraction.}
Given the binary under test, \ourtool{} reconstructs the inter-procedural control-flow graph (ICFG) by static disassembly. The extraction recovers basic-block boundaries and branch targets within each function to
produce intra-procedural CFGs, and identifies call sites and their callees to build the inter-procedural call graph. Direct calls and statically determinable branches are resolved at this stage with high accuracy, as are the recurring MIPS PIC dispatch patterns whose callee the disassembler can recover. The remaining indirect transfers are left to the dynamic component, since even a partial edge set gives a useful initial estimate.

\paragraph{Initial distance scores computation.}
With the ICFG in hand, \ourtool{} runs a shortest-path pass over the linked CFG-plus-call-graph to assign each basic block a distance score, taken as the harmonic mean of its weighted shortest-path distances to the target blocks. The DBB classification pass
(Section~\ref{sec:dbb-classifier}) runs as a postprocessing step over the same graph, labeling loop-gating and chokepoint branches. Both the distance table and the DBB annotations are serialised to a compact binary file that the fuzzer loads into shared memory
at startup, making the distance model available to the power scheduler from the very first execution.

\paragraph{Runtime graph refinement.}
Indirect transfer points left unresolved by static analysis introduce blind spots: the initial distance computation may underestimate, or even miss entirely, paths that route through
function pointers or computed jumps. To close these gaps, \ourtool{} instruments every indirect branch and call in the QEMU translation layer. The first time an indirect transfer is observed to reach a new destination address, the enclosing (caller, callee) edge is added to a live call-graph replica maintained outside the emulation environment. An out-of-process watcher, signalled by the fuzzer, reruns the distance computation incorporating the new edges, and updates the shared-memory distance table used by the running fuzzer. Over a campaign the live graph accumulates the indirect edges the fuzzer exercises, progressively reducing the blind spots introduced by purely static analysis.

\medskip

These stages are conceptually straightforward but hide several crucial details; Section~\ref{sec:implementation} covers the implementation aspects that matter most.

\subsection{Three-Condition DBB Classification}
\label{sec:dbb-classifier}

\windranger{} introduced DBBs as a noise-reduction mechanism for directed fuzzing: instead of averaging distances over all executed basic blocks, the fuzzer restricts its distance computation to blocks where branch direction actually determines reachability to the target.

\windranger{}'s classifier defines a DBB as a block $b$ such that (i)~$b$ can reach the target, and (ii)~at least one successor of $b$ cannot reach the target. In what follows, we refer to this second constraint as Condition~2a, following the labels of
Table~\ref{tab:dbb-patterns}.
As shown in Section~\ref{sec:mot:dbb}, Condition~2a is formulated purely in terms of reachability over the full control-flow graph $G$. For a block $b$ inside a loop with a back-edge successor $b_\text{back}$ and a forward-edge successor $b_\text{fwd}$, both successors can eventually reach the target via some path in $G$ (either by looping around again, or by direct forward progress). The condition
$\exists\, s \in \mathrm{succ}(b) \colon \lnot\,\text{reach}(s, t)$ therefore fails, and $b$ is not classified as a DBB. The underlying deficiency is that reachability over $G$ is path-insensitive with respect to back edges: it cannot encode the distinction between $b_\text{back}$ reaching the target only by re-entering the loop and $b_\text{fwd}$ reaching it by direct forward progress.

\paragraph{The three-condition definition.}
\ourtool{}'s classification retains the two conditions from \windranger{} but refines the second one by adding two new variants that reason about the
acyclic graph $G_\text{fwd}$ obtained by removing the back edges of a depth-first traversal of $G$. Formally, the classification exploits two distinct reachability relations:

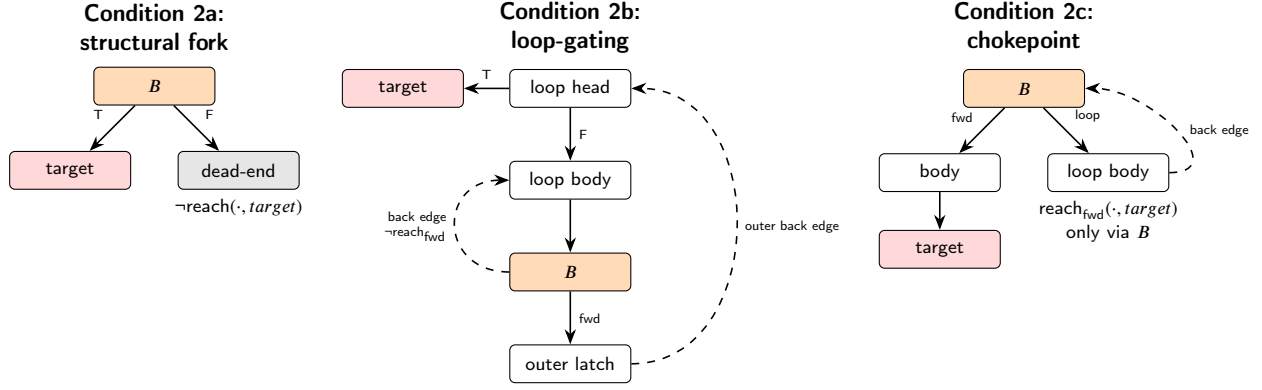
\begin{figure*}[t]
\centering
\begin{tikzpicture}[
  bb/.style={draw, rounded corners=2pt, fill=white, font=\scriptsize,
             minimum width=1.6cm, minimum height=5mm, inner sep=2pt, align=center},
  tgt/.style={draw, rounded corners=2pt, fill=red!15, font=\scriptsize,
              minimum width=1.6cm, minimum height=5mm, inner sep=2pt, align=center},
  dead/.style={draw, rounded corners=2pt, fill=gray!20, font=\scriptsize,
               minimum width=1.6cm, minimum height=5mm, inner sep=2pt, align=center},
  dbb/.style={draw, rounded corners=2pt, fill=orange!28, font=\scriptsize,
              minimum width=1.6cm, minimum height=5mm, inner sep=2pt, align=center},
  arr/.style={->, >=Stealth, semithick},
  darr/.style={->, >=Stealth, semithick, dashed},
  lbl/.style={font=\small\bfseries, align=center},
  ann/.style={font=\scriptsize, text=black, align=center},
]
\begin{scope}[xshift=-1.5cm]
  \node[lbl] (la) at (0,0) {Condition 2a:\\structural fork};
  \node[dbb,  below=1mm of la]                (B2a)  {$B$};
  \node[tgt,  below left=6mm and -5mm of B2a] (T2a)  {target};
  \node[dead, below right=6mm and -5mm of B2a](D2a)  {dead-end};
  \draw[arr] (B2a) -- node[above left, font=\tiny]{T} (T2a);
  \draw[arr] (B2a) -- node[above right,font=\tiny]{F} (D2a);
  \node[ann, below=0mm of D2a] {$\neg\text{reach}(\cdot,target)$};
\end{scope}

\begin{scope}[xshift=4cm]
  \node[lbl] (lb) at (0,0) {Condition 2b:\\loop-gating};
  \node[bb,  below=1mm of lb]    (H2b)  {loop head};
  \node[tgt, left=6mm of H2b]    (T2b)  {target};
  \node[bb,  below=7mm of H2b]   (IB2b) {loop body};
  \node[dbb, below=7mm of IB2b]  (B2b)  {$B$};
  \node[bb,  below=7mm of B2b]   (OL2b) {outer latch};
  \draw[arr] (H2b) -- node[above,font=\tiny]{T} (T2b);
  \draw[arr] (H2b) -- node[right,font=\tiny]{F} (IB2b);
  \draw[arr] (IB2b) -- (B2b);
  \draw[darr](B2b.west) to[out=180, in=180, looseness=2.0]
    node[left,font=\tiny,align=right]{back edge\\$\neg\text{reach}_\text{fwd}$} (IB2b.west);
  \draw[arr] (B2b) -- node[right,font=\tiny]{fwd} (OL2b);
  \draw[darr](OL2b.east) to[out=0, in=0, looseness=1.3]
    node[right,font=\tiny,align=left]{outer back edge} (H2b.east);
\end{scope}

\begin{scope}[xshift=10cm]
  \node[lbl] (lc) at (0,0) {Condition 2c:\\chokepoint};
  \node[dbb,  below=1mm of lc]                (B2c)  {$B$};
  \node[bb,   below left=6mm and -5mm of B2c] (V2c)  {body};
  \node[tgt,  below=5mm of V2c]               (T2c)  {target};
  \node[bb,   below right=6mm and -5mm of B2c](E2c)  {loop body};
  \draw[arr] (B2c) -- node[above left, font=\tiny]{fwd} (V2c);
  \draw[arr] (V2c) -- (T2c);
  \draw[arr] (B2c) -- node[above right,font=\tiny]{loop} (E2c);
  \draw[darr](E2c.east) to[out=0, in=0, looseness=1.4]
    node[right,font=\tiny,align=left]{~back edge} (B2c.east);
  \node[ann, below=0mm of E2c] {$\text{reach}_\text{fwd}(\cdot,target)$\\[1pt]only via $B$};
\end{scope}
\end{tikzpicture}
\caption{The three DBB conditions in \ourtool{}. Block $B$ is always the candidate DBB (orange). \textbf{Condition 2a}: one successor has no path to target at all. \textbf{Condition 2b}: $B$ is a loop latch whose back-edge successor cannot reach the target in $G_\text{fwd}$; the target sits at the loop top, so the forward successor reaches it only by traversing the outer back edge. \textbf{Condition 2c}: $B$'s successors split, so that one reaches the target forward while the other reaches it only by looping back and re-entering $B$.}
\label{fig:three-conditions}
\end{figure*}

\begin{itemize}
  \item $\text{reach}(u, t)$: $t$ is reachable from $u$ in $G$ (including via back edges).
  \item $\text{reach}_\text{fwd}(u, t)$: $t$ is reachable from $u$ in $G_\text{fwd}$.
\end{itemize}

\begin{definition}
A block $b$ with successors $S(b)$ is a DBB with respect to target set $T$ if and only if $b$ has at
least two successors, the set
$\mathcal{T} = \{\, t \in T : \text{reach}(b, t) \wedge \exists\, s \in S(b) : \text{reach}(s, t)\,\}$
of targets reachable through a successor (not just from $b$ directly) is non-empty, and, quantifying over $\mathcal{T}$, at least one of the following holds:

\begin{itemize}

\item \textbf{Condition 2a (structural fork):}
$\exists\, s \in S(b), t \in \mathcal{T} : \lnot\,\text{reach}(s, t)$. One successor loses a target that $b$ can still reach; taking that branch is a definitive divergence.

\item \textbf{Condition 2b (loop-gating DBB):}
The successor set $S(b)$ has a non-empty back-edge partition $B(b)$ and a non-empty forward
partition $F(b)$; and
$\forall\, b' \in B(b), t \in \mathcal{T} : \lnot\,\text{reach}_\text{fwd}(b', t)$.
No back-edge successor can reach any target in $G_\text{fwd}$: the only way to reach the
target via the back edge is to loop around through $b$ again.  Some forward-edge successor
$s \in F(b)$ still reaches a target in $G$ ($\exists\, t \in \mathcal{T} : \text{reach}(s, t)$).

\item \textbf{Condition 2c (mandatory re-entry chokepoint):} $b$ is not classified by Conditions~2a or~2b. For some target $t \in \mathcal{T}$, one successor $s \in S(b)$ still reaches $t$ in $G_\text{fwd} \setminus \{b\}$ while another successor $s' \in S(b)$ does not, so $b$ dominates $t$ on every forward path from $s'$. Thus $s$ makes forward progress around $b$, while $s'$ reaches $t$ only by looping back and re-entering $b$.

\end{itemize}

\end{definition}

Figure~\ref{fig:three-conditions} illustrates the three conditions on minimal abstract CFGs, clarifying the structural patterns each one targets.
The three conditions are disjoint by construction: Condition~2a is evaluated first and short-circuits; Condition~2b fires on the forward branch when no back-edge successor reaches a target in $G_\text{fwd}$; Condition~2c is then evaluated only on the blocks that 2a and 2b leave unclassified, testing whether deleting $b$ from $G_\text{fwd}$ severs one successor's route to a target while another successor still reaches it, so that $b$ is a forward articulation point. Conditions~2a and~2b cost $O(|S(b)| \cdot |\mathcal{T}|)$ reachability queries, memoised on $G$; Condition~2c additionally builds $G_\text{fwd} \setminus \{b\}$ once per candidate and tests forward reachability from each successor on it, which is the dominant term of the classification pass.

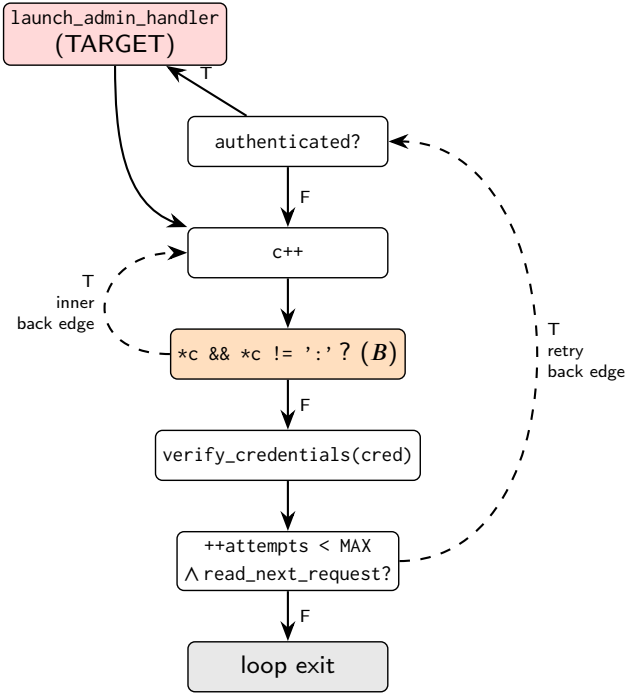
\begin{figure}
\centering
\resizebox{\columnwidth}{!}{%
\begin{tikzpicture}[
  bb/.style={draw, rounded corners=2pt, fill=white, font=\scriptsize,
             minimum width=2.0cm, minimum height=5mm, inner sep=2pt, align=center},
  tgt/.style={draw, rounded corners=2pt, fill=red!15, font=\scriptsize,
              minimum width=2.0cm, minimum height=5mm, inner sep=2pt, align=center},
  dbb/.style={draw, rounded corners=2pt, fill=orange!25, font=\scriptsize,
              minimum width=2.0cm, minimum height=5mm, inner sep=2pt, align=center},
  exit/.style={draw, rounded corners=2pt, fill=gray!18, font=\scriptsize,
               minimum width=2.0cm, minimum height=5mm, inner sep=2pt, align=center},
  arr/.style={->, >=Stealth, semithick},
  darr/.style={->, >=Stealth, semithick, dashed},
  ann/.style={font=\tiny, align=center},
]
  \node[bb]  (au)  {\texttt{authenticated?}};
  \node[tgt, above left=5mm and -4mm of au]  (tg)  {\texttt{launch\_admin\_handler}\\(TARGET)};
  \node[bb,  below=6mm of au]  (sb)  {\texttt{c++}};
  \node[dbb, below=5mm of sb]  (sl)  {\texttt{*c \&\& *c != ':'}\,? ($B$)};
  \node[bb,  below=5mm of sl]  (ve)  {\texttt{verify\_credentials(cred)}};
  \node[bb,  below=5mm of ve]  (dw)  {\texttt{++attempts < MAX}\\$\wedge$\,\texttt{read\_next\_request?}};
  \node[exit, below=5mm of dw]  (ex) {loop exit};

  \draw[arr] (au) -- node[above, font=\tiny]{T} (tg);
  \draw[arr] (au) -- node[right, font=\tiny]{F} (sb);
  \draw[arr] (tg) to[out=-90, in=160] (sb.north west);
  \draw[arr] (sb) -- (sl);
  \draw[darr](sl.west) to[out=180, in=180, looseness=2.4]
    node[left, font=\tiny, align=right]{T\\inner\\back edge} (sb.west);
  \draw[arr] (sl) -- node[right, font=\tiny]{F} (ve);
  \draw[arr] (ve) -- (dw);
  \draw[darr](dw.east) to[out=0, in=0, looseness=1.15]
    node[right, font=\tiny, align=left]{T\\retry\\back edge} (au.east);
  \draw[arr] (dw) -- node[right, font=\tiny]{F} (ex);
\end{tikzpicture}%
}
\caption{Condition~2b on the CFG of the listing from Pattern~2.  The target
  \texttt{launch\_admin\_handler} is dispatched at the loop top; the inner credential-scan latch
  $B$ (\texttt{*c != ':'}, orange) sits below it.  $B$'s back-edge successor cannot reach the
  target in $G_\text{fwd}$, and its forward successor reaches the target only by traversing the
  outer retry back edge, never forward.  Condition~2b therefore classifies $B$ as a DBB, which
  \windranger{} misses.}
\label{fig:cond2b-loop}
\end{figure}

\paragraph{Illustrating Condition 2b.}
Returning to the listing from Pattern~2 (bounded-retry authentication gate), the privileged handler \texttt{launch\_admin\_handler} (the target~$t$) is dispatched at the \emph{top} of the retry loop, guarded by the \texttt{authenticated} flag. The candidate is the inner credential-scan latch $B$ (\texttt{*c \&\& *c != ':'}), which sits below the target inside the loop. Figure~\ref{fig:cond2b-loop} traces the Condition~2b check. $B$ has one back-edge successor (the scan body \texttt{c++}) and one forward successor (\texttt{verify\_credentials},
continuing to the outer retry test). In $G_\text{fwd}$ neither successor reaches the target, because the target sits at the loop top and every route to it must cross a back edge: the
back-edge successor fails $\text{reach}_\text{fwd}(\cdot, target)$, while the forward successor still reaches the target in the full graph $G$ by traversing the outer retry back edge. Condition~2b therefore fires and $B$ is classified as a DBB. \windranger{}'s structural check
does not classify it, since both successors are reachable-to-target in $G$, as demonstrated in
Section~\ref{sec:mot:dbb}.

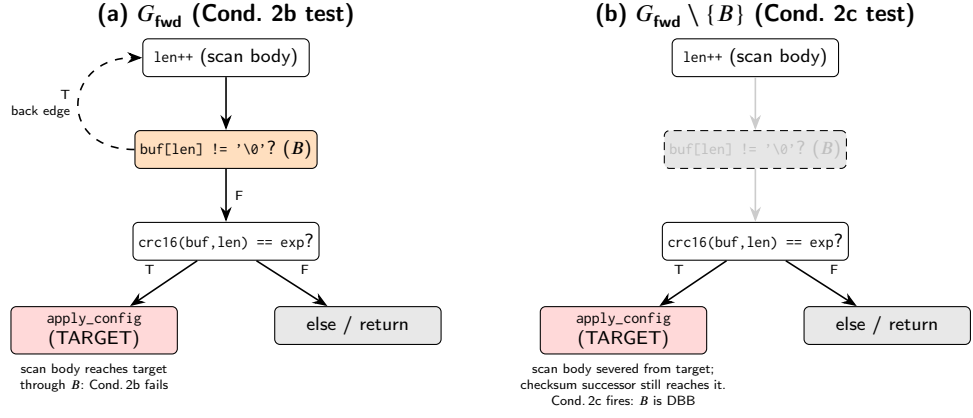
\begin{figure*}[t]
\centering
\begin{tikzpicture}[
  bb/.style={draw, rounded corners=2pt, fill=white, font=\scriptsize,
             minimum width=2.2cm, minimum height=5mm, inner sep=2pt, align=center},
  tgt/.style={draw, rounded corners=2pt, fill=red!15, font=\scriptsize,
              minimum width=2.2cm, minimum height=5mm, inner sep=2pt, align=center},
  dbb/.style={draw, rounded corners=2pt, fill=orange!25, font=\scriptsize,
              minimum width=2.2cm, minimum height=5mm, inner sep=2pt, align=center},
  removed/.style={draw, rounded corners=2pt, fill=gray!20, font=\scriptsize,
                  minimum width=2.2cm, minimum height=5mm, inner sep=2pt,
                  align=center, densely dashed, text=gray!50},
  exit/.style={draw, rounded corners=2pt, fill=gray!18, font=\scriptsize,
               minimum width=2.2cm, minimum height=5mm, inner sep=2pt, align=center},
  arr/.style={->, >=Stealth, semithick},
  darr/.style={->, >=Stealth, semithick, dashed},
  garr/.style={->, >=Stealth, semithick, gray!40},
  lbl/.style={font=\small\bfseries},
]
\begin{scope}[xshift=0cm]
  \node[lbl] at (0,0) {(a) $G_\text{fwd}$ (Cond.\ 2b test)};
  \node[bb,   below=3mm]                     (SB) {\texttt{len++} (scan body)};
  \node[dbb,  below=7mm of SB]               (SL) {\texttt{buf[len] != '\textbackslash0'}? ($B$)};
  \node[bb,   below=7mm of SL]               (CK) {\texttt{crc16(buf,len) == exp}?};
  \node[tgt,  below left=6mm and -6mm of CK] (T)  {\texttt{apply\_config}\\(TARGET)};
  \node[exit, below right=6mm and -6mm of CK](EX) {else / return};
  \draw[arr]  (SB) -- (SL);
  \draw[darr] (SL.west) to[out=180, in=180, looseness=2.2]
    node[left, font=\tiny, align=right]{T\\back edge} (SB.west);
  \draw[arr]  (SL) -- node[right, font=\tiny]{F} (CK);
  \draw[arr]  (CK) -- node[above left,  font=\tiny]{T} (T);
  \draw[arr]  (CK) -- node[above right, font=\tiny]{F} (EX);
  \node[font=\tiny, text=black, below=0mm of T, align=center]
    {scan body reaches target\\through $B$: Cond.\,2b fails};
\end{scope}

\begin{scope}[xshift=7cm]
  \node[lbl] at (0,0) {(b) $G_\text{fwd}\setminus\{B\}$ (Cond.\ 2c test)};
  \node[bb,      below=3mm]                      (SBx){\texttt{len++} (scan body)};
  \node[removed, below=7mm of SBx]               (SLx){\texttt{buf[len] != '\textbackslash0'}? ($B$)};
  \node[bb,      below=7mm of SLx]               (CKx){\texttt{crc16(buf,len) == exp}?};
  \node[tgt,     below left=6mm and -6mm of CKx] (Tx) {\texttt{apply\_config}\\(TARGET)};
  \node[exit,    below right=6mm and -6mm of CKx](EXx){else / return};
  \draw[garr] (SBx) -- (SLx);
  \draw[garr] (SLx) -- (CKx);
  \draw[arr]  (CKx) -- node[above left,  font=\tiny]{T} (Tx);
  \draw[arr]  (CKx) -- node[above right, font=\tiny]{F} (EXx);
  \node[font=\tiny, text=black, below=0mm of Tx, align=center]
    {scan body severed from target;\\checksum successor still reaches it.\\Cond.\,2c fires: $B$ is DBB};
\end{scope}
\end{tikzpicture}
\caption{Two-step Condition~2c test on the CFG of the listing from Pattern~3 (checksum-gated
  configuration copy).  \textbf{(a)} In $G_\text{fwd}$, the scan latch $B$'s back-edge successor
  (\texttt{len++}) reaches the target only through $B$, while its forward successor
  (\texttt{crc16}) reaches it without $B$, so Condition~2b does not fire.
  \textbf{(b)} Removing $B$ from $G_\text{fwd}$ (shown greyed) severs the back-edge successor from
  the target while the forward successor still reaches it, so $B$ is a forward articulation point
  and Condition~2c classifies it as a DBB.}
\label{fig:cond2c-scan}
\end{figure*}

\paragraph{Illustrating Condition 2c.}
In the listing from Pattern~3 (checksum-gated configuration copy), the scan latch $B$ (\texttt{buf[len] != '\textbackslash0'}) has a back-edge successor (\texttt{len++}, the scan body) and a forward successor (the checksum test \texttt{crc16}). Figure~\ref{fig:cond2c-scan} traces
the two-step Condition~2c test. In $G_\text{fwd}$ the scan body still reaches the target, but only \emph{through} $B$ (\texttt{len++} $\to B \to$ \texttt{crc16} $\to$ \texttt{apply\_config}), so
Condition~2b does not fire. Removing $B$ from $G_\text{fwd}$ severs that route: the scan body can no longer reach the target, while the forward successor still reaches it through the checksum test. Because deleting $B$ disconnects one successor from the target while another bypasses it, $B$ is a forward articulation point, and Condition~2c classifies it as a DBB.  The same structure arises in the listing from Pattern~1 (protocol-header enumeration), where the branch block's false edge re-enters the loop head, which reaches the target only by looping back through that branch block. Because the articulation-point test ranges over \emph{all} successors of $b$ rather than only its back-edge successors, Condition~2c captures such a forward branch at a loop \emph{head} and at a latch.

\paragraph{DBB explosion guards.}
The extended classification introduces more DBBs than \windranger{}, raising a potential concern about signal dilution. Two properties prevent over-classification: (i)~Condition~2b requires \emph{all} back-edge successors to fail the forward-reachability test; a node with even one independent forward path is not classified as a loop-gating DBB;
(ii)~Condition~2c requires a whole-graph reachability property in $G_\text{fwd}$, namely that deleting $b$ severs one successor's route to the target, not just a local branch property.

\subsection{Satisfiability-Adjusted Distance}
\label{sec:sat-score}\label{sec:distance}

A DBB identifies a block as a decision point; it does not quantify how far a given seed is from satisfying the block's branch predicate. This quantification is provided in \ourtool{} by the \emph{branch satisfiability score}.

\begin{definition}
For a DBB $i$ traversed by seed $q$, the satisfiability score $\sigma_i(q) \in [0, 1]$ measures
the degree to which the comparison operands observed at $i$ during the execution of $q$ match the
values that would direct execution toward the target.
\end{definition}

\ourtool{} computes the satisfiability score from two sources of runtime information.

\paragraph{Branch-direction bits.}
QEMU's MIPS front end materialises each conditional branch outcome in a TCG global (\texttt{bcond}), set to~1 when the branch is taken. An instrumentation helper invoked during branch translation records that outcome alongside both successor addresses, which are checked against the preloaded distance table to determine which direction leads toward the target. This populates two shared-memory arrays: \texttt{target\_cond\_bits[i]}, the direction required to reach the target at DBB~$i$, and \texttt{cond\_bits[i]}, the direction actually taken, or zero if DBB~$i$ was not reached during the current execution. If \texttt{cond\_bits[i] == target\_cond\_bits[i]}, the seed already takes the correct branch at DBB~$i$ and $\sigma_i(q)$ is set to~1.0 immediately.

\paragraph{Operand-level similarity.}
When the branch direction does not match the target, the satisfiability score is computed from the comparison operands logged by CmpLog instrumentation~\cite{cmplog}. CmpLog intercepts comparison operations during TCG translation and records the operand pair $(\ell, r)$ for each executed comparison.

The score depends on the kind of comparison, which \ourtool{} reads from the CmpLog comparison attribute. For equality and inequality comparisons ($=$, $\neq$) and for floating-point comparisons, the score is the fraction of bits that agree between the two operands:
\[
  \sigma_\text{bit}(\ell, r, w)
    = \frac{\texttt{popcount}(\lnot(\ell \oplus r) \mathbin{\&} m_w)}{w},
\]
where $w \in \{8, 16, 32, 64\}$ is the comparison width and $m_w = 2^w - 1$ is the corresponding bit mask; an exact match short-circuits to $1.0$.  Floating-point operands use the same bitwise measure because their raw encodings cannot be differenced as integers. The score is sensitive to small perturbations: as $\ell$ approaches $r$ bit by bit, it increases toward~$1$.

For ordered comparisons ($<$, $>$, $\le$, $\ge$), a bitwise measure is uninformative, since two numerically adjacent values may differ in many bits. \ourtool{} instead scores these by numeric proximity to the satisfying side. If the predicate is already satisfied by the observed operands the score is~$1.0$; otherwise it is
\[
  \sigma_\text{ord}(\ell, r, w)
    = \max\!\left(0,\; 1 - \frac{\log_2\!\bigl(|\ell - r| + 2\bigr)}{w}\right),
\]
where $|\ell - r|$ is the magnitude by which the predicate fails. This formulation makes the branch direction intrinsic: operands on the wrong side of a strict boundary never score~$1.0$, and the score grows as the operand moves toward the threshold.

For string comparisons (intercepted via \texttt{strcmp}, \texttt{strncmp}, \texttt{memcmp}, and their variants) the metric counts the fraction of byte positions that agree within the shorter operand:
\[
  \sigma_\text{str}(a, b, n)
    = \frac{|\{ i < n : a[i] = b[i] \}|}{n}
\]
where $n = \min(|a|, |b|)$.  Normalising by the shorter length avoids penalising a seed whose string is a correct prefix but shorter than the expected operand.

When a seed executes a DBB multiple times in one run (as in a loop), the satisfiability score is the maximum over all CmpLog hits, capturing the best-case progress at that DBB. Where no comparison is logged, $\sigma_i = 0$ and $d_i^\star$ falls back to the structural distance.

\paragraph{From difficulty weighting to progress weighting.}
\windranger{} scales each DBB's structural distance by $\Psi_i$, a coarse estimate of how hard that DBB's predicate is to penetrate, derived from the number of input bytes that affect it. The weight describes the predicate, not the seed: two seeds reaching the same DBB with different operand values
are scored alike. \ourtool{} replaces it with an adjustment driven by how close the seed itself came to satisfying the branch.

\begin{definition}
For DBB~$i$ hit by seed $q$, the satisfiability-adjusted per-DBB distance is:
\[
  d_i^\star(q) = \frac{d_i}{1 + 99 \cdot \sigma_i(q)}
\]
where $d_i$ is the static distance of DBB~$i$ from Section~\ref{sec:binary-pipeline}.
\end{definition}

At $\sigma = 0$ the adjusted distance equals the raw structural distance; at $\sigma = 1$ it is reduced by a factor of~100. The factor~100 is chosen so that partial progress registers early, with $\sigma = 0.01$ already halving the adjusted distance, while the remaining range compresses gradually so that satisfiability never overrides structural distance outright: a seed with a perfect satisfiability score at a distant DBB does not necessarily rank better than a seed with a moderate satisfiability score at a structurally proximate DBB.

\paragraph{Reciprocal-sum aggregation.}
A seed typically hits multiple DBBs in a single execution. \ourtool{} combines their satisfiability-adjusted per-DBB distances into a single per-seed distance as the reciprocal of the summed reciprocals:
\[
  D(q) = \left(\;\sum_{i \,:\, \text{DBB}_i \text{ hit by } q}
    \frac{1}{d_i^\star(q)}\right)^{-1}.
\]
Like the harmonic mean, this combiner is dominated by its smallest term: if a seed reaches any single DBB with high satisfiability score and low structural distance, the aggregate reflects that progress regardless of how far the other DBBs remain. By contrast, the arithmetic mean used in \aflgo{} and \windranger{} is pulled up by the largest distance, making it insensitive to near-solutions at individual DBBs. As a running example, a seed that hits three DBBs with satisfiability-adjusted distances $\{100,\; 5,\; 200\}$ receives an aggregate of $\bigl(\tfrac{1}{100}+\tfrac{1}{5}+\tfrac{1}{200}\bigr)^{-1} \approx 4.65$, reflecting the near-target DBB, whereas the arithmetic mean would report $\approx 101.7$, i.e.\ a seed that appears far from the target. Seeds that reach a target basic block are assigned distance~$0.0$ and are always ranked first.

\subsection{Power Scheduling and Multi-Target Fairness}
\label{sec:scheduling}\label{sec:stagnation}

\ourtool{} retains \aflgo{}'s simulated annealing framework: a global temperature $T(t) \in [0, 1]$ parameterised by $p = t / t_x$ (elapsed time divided by the annealing duration). Four cooling schedules are available (exponential, logarithmic, linear, quadratic); the logarithmic schedule cools fastest, the exponential $T = 1 / 20^p$ is the default, and all four reach $T = 0.05$ at $t = t_x$.  At $T = 1$ all seeds receive equal energy; at $T = 0$ energy is concentrated on seeds closest to the target.

The base performance score is adjusted by a distance-based power factor following the \aflgo{}
formulation:
\begin{align*}
  p_f &= 2^{\,2\log_2(F_\text{max}) \cdot (p_\text{norm} - 0.5)} \\
  p_\text{norm} &= (1 - \hat{D})(1 - T) + 0.5\, T
\end{align*}
where $\hat{D} \in [0,1]$ is the normalised distance of the seed relative to current corpus
extremes and $F_\text{max} = 32$ is the maximum power factor, which bounds the energy a seed can gain or lose.

\paragraph{Per-DBB independent annealing.}
\ourtool{}'s per-DBB annealing addresses the discovery-time inequality identified in Section~\ref{sec:mot:annealing}. Each DBB~$i$ maintains its own temperature $T_i \in [0, 1]$ governed by an independent cooling clock that starts at the moment DBB~$i$ is first hit:
\[
  T_i(t) = \mathcal{C}\!\left(\frac{t - t_i^\text{first}}{t_x \cdot 60000}\right)
\]
where $t_i^\text{first}$ is the timestamp (in milliseconds) at which DBB~$i$ was first reached, the denominator is the annealing duration expressed in milliseconds, and $\mathcal{C}(\cdot)$ is the same cooling schedule selected for the global temperature (exponential by default, or one of the logarithmic, linear, and quadratic alternatives). The temperature is floored at~0.001 when active (preventing numerical underflow) and frozen at~0.0 once DBB~$i$ has been solved (both branch directions observed at that DBB, so its outcome is no longer input-controlled from the fuzzer's viewpoint).

Per-DBB temperatures enter the power schedule via a geometric mean over all DBBs hit by a seed:
\begin{align*}
    c_\text{per-DBB}(q)
      &= \left(\prod_{i=1}^n p_{f,i}\right)^{1/n} \\
    p_{f,i} &= 2^{\,2\log_2(F_\text{max}) \cdot (p_i - 0.5)} \\
    p_i     &= (1 - \hat{d}_i)(1 - T_i) + 0.5\, T_i \\
    \hat{d}_i &= \max\!\left(\tilde{d}_i,\; 1 - \sigma_i\right)
  \end{align*}
The geometric mean keeps the adjustment independent of how many DBBs a seed hits, so that one extreme DBB cannot dominate the others. Here $\tilde{d}_i \in [0,1]$ is the per-DBB static distance rescaled to the static range of all DBB distances, and $\hat{d}_i$ is the \emph{worse} of that structural distance and the unsatisfiability $1 - \sigma_i$: a DBB pulls energy only when it is both structurally close \emph{and} nearly satisfied, mirroring the global factor's use of $1 - \hat{D}$.

Global and per-DBB temperatures interact: while the global schedule drives the campaign from exploration toward exploitation, a DBB first hit late still opens with a high $T_i$, granting it a local exploration burst that counteracts the global exploitation trend.

\paragraph{Satisfiability-guided corpus management.}
Consistent with the satisfiability-adjusted distance, the corpus management mechanisms are extended
to account for satisfiability progress.

Standard AFL admits a new seed to the queue only if it exercises previously unseen edges. \ourtool{} additionally admits a seed that improves the satisfiability score at any DBB relative to the current corpus best, even if no new edges are covered. This extended notion of \emph{interestingness}
ensures that incremental progress toward satisfying a hard comparison predicate is preserved in the corpus rather than discarded.

The culling routine is similarly extended. For each DBB, \ourtool{} maintains two favoured seeds: the one closest to the target under the satisfiability-adjusted distance $d_i / (1 + 99\sigma_i)$, and the one achieving the highest satisfiability score at that DBB's predicate. Both are protected from culling while the DBB is unsolved. Once the DBB is solved, the satisfiability leader is released, since satisfiability progress is irrelevant once the branch is flippable in both directions, whereas the distance leader is retained, since proximity to the target remains meaningful after the branch is passable. Seeds admitted solely for satisfiability improvement, contributing no new coverage, are then evicted, preventing the corpus from accumulating stale seeds that no longer serve a purpose.

\paragraph{Relative-saturation energy decay.}
\ourtool{} tracks, for each target~$i$, the number of executions that reach it, $\mathit{reach}_i$, and the wall-clock time of the most recent saved input reaching it. A target's \emph{saturation} combines a relative-frequency term with a stagnation valve:
\[
  s_i = \max\!\left(\;
    \underbrace{\frac{\mathit{reach}_i}{\max_j \mathit{reach}_j}}_{\text{relative frequency}},\;\;
    \underbrace{\frac{a_i}{a_i + \tau}}_{\text{stagnation valve}}
  \right),
\]
where $a_i$ is the time elapsed since the last saved input reaching target~$i$ was last reached and $\tau$ is a stagnation window of 30~minutes; the relative-frequency term is active only when more than one target has been reached. Each target is assigned an energy weight $w_i = f + (1 - f)(1 - s_i)$ that decays from~$1$ toward a floor~$f = 0.25$ as the target saturates. The weight sets an upper bound on a seed's energy: a seed's havoc budget is capped at $w \cdot c_\text{max}$, never below the baseline score, so a seed already below the ceiling is left untouched and only over-energetic seeds tied to a saturated target are clipped. The weight acts through a second lever as well, raising the probability that a seed bound to a saturated target is skipped in a given cycle; both levers drain energy in the same direction. The governing weight is the largest among the targets the seed reaches (so a seed is judged by its most under-served target), or the global minimum weight if it reaches no target. A target that is either well-reached \emph{or} long-stagnant therefore lowers the ceiling of the seeds bound to it, whereas an under-served or recently productive target keeps it high. The stagnation valve is what allows energy to drain away from a target that is reached but no longer productive, without a crash ever being observed there, distinguishing \ourtool{} from saturation schemes that require a bug signal.

\paragraph{Target-fairness favouring.}
Complementary to the continuous decay, the corpus-favouring routine biases selection toward seeds that reach an under-reached target relative to the most-reached one. This is self-balancing and target-agnostic: as a starved target accumulates reaching seeds it stops being favoured, and energy shifts to whichever target is now the most neglected.

\paragraph{Coverage-stall reheating.}
Directed pressure can trap the campaign in a local optimum where no new coverage appears for a long period. Once the time since the last new path exceeds a stall window of 10 minutes, \ourtool{} temporarily \emph{reheats} the corpus, reviving seeds previously culled as unproductive in case one unlocks a new edge toward a target. Reheating preserves the distance bias (seeds are still ordered by proximity) and clears automatically as soon as a new path is found.

\begin{table*}[t]
  \centering
  \caption{Design comparison of \aflgo{}, \windranger{}, and \ourtool{} across all four problem dimensions.}
  \label{tab:comparison}
  \resizebox{\linewidth}{!}{%
  \begin{tabular}{llccc}
  \toprule
  Problem & Dimension & \aflgo{} & \windranger{} & \ourtool{} \\
  \midrule
  P1: Binary analysis
    & Target type         & Source required & Source required & Closed-source binary \\
    & Call-graph edges    & Static (LLVM)   & Static (LLVM)   & Static (IDA) + runtime refinement \\
  \midrule
  P2: DBB classification
    & DBB concept         & None                   & Reachability-based (2a)          & Loop-aware (2a+2b+2c) \\
    & Loop patterns       & Not handled            & Not handled               & Handled via 2b and 2c \\
  \midrule
  P3: Distance metric
    & Distance over       & All BBs                & DBBs only                 & DBBs only \\
    & Satisfiability signal & None                 & Predicate difficulty $\Psi_i$ & Continuous $\sigma_i \in [0,1]$ \\
    & Signal semantics    & None                    & Static per-predicate cost & Per-seed progress toward the branch \\
    & Adjustment formula  & None                   & $d_i \cdot \Psi_i$        & $d_i / (1 + 99\sigma_i)$ \\
    & Aggregation         & Arithmetic mean        & Arithmetic mean           & Reciprocal-sum \\
  \midrule
  P4: Power scheduling
    & Annealing           & Global only            & Global only               & Global + per-DBB \\
    & Corpus admission    & Edge coverage          & Edge coverage             & Edge coverage + satisfiability progress \\
    & Target fairness & None            & None     & Saturation decay + fairness favouring \\
    & Coverage Stall & None & None & Corpus reheating \\
  \bottomrule
  \end{tabular}%
  }
\end{table*}

\subsection{Discussion}
\label{sec:discussion}

Table~\ref{tab:comparison} puts \ourtool{}'s key contributions in perspective against \aflgo{} and \windranger{} across all design dimensions. The four contributions are largely orthogonal, and the table makes explicit that \aflgo{} and \windranger{} share the same design for P1 (both require source code) and for P4 (both driven by a single global clock), while \windranger{} improves on \aflgo{} for P2 and P3 but only partially, stopping at Condition~2a and at a seed-independent difficulty weight, respectively. \ourtool{} extends both of these dimensions and additionally addresses P1 and P4.

\section{Implementation}
\label{sec:implementation}

We realise \ourtool{}'s design on top of \firmafl{}~\cite{firmafl} and its MIPS-targeted QEMU builds. This section covers the techniques and optimisations required to apply DGF to closed-source firmware binaries and to operate effectively within the QEMU emulation layer. 

\subsection{Binary-Level Graph Extraction}
\label{sec:impl:extraction}

The static analysis phase of DGF requires three representations of the binary under test: intra-procedural
control-flow graphs (CFGs), an inter-procedural call graph (CG), and a call-site map that links
each call instruction to its callee. In source-available settings these are obtained from
LLVM~IR during compilation. For closed-source firmware, \ourtool{} extracts them from the disassembled binary using a scripted IDA~Pro~\cite{ida} pass.

The script traverses every function in the binary and records, for each basic block,
the block's start and end addresses and the set of successor addresses derived from the block-terminator instruction. Every call site whose callee can be resolved is recorded in the call-site map with the absolute address of the call instruction and the address or symbol name of the callee. The inter-procedural CG is extracted separately via IDA's built-in call-chart generator, which emits the caller-callee relationships as a graph description file.

A parser stage ingests these raw records and synthesises the intra-procedural CFGs as directed
graphs, adding explicit edges for fallthrough successors where IDA does not emit a branch instruction. The output graphs and index files match what the \aflgo{} distance-computation scripts~\cite{dgf1} expect, allowing \ourtool{} to reuse that pipeline with minimal modification.

\paragraph{Address scheme.}
DGF frameworks that operate at the source level identify basic blocks by their file-and-line number, which is stable across recompilations. \ourtool{} identifies blocks by their offset
relative to the binary's load base, which is invariant under address-space layout randomisation
and across different loader configurations. At runtime, the offset is converted to an absolute address by adding the observed load base, which QEMU reports when mapping the binary into the emulated address space.

\paragraph{MIPS PIC handling.}
On MIPS, the position-independent calling convention reserves register \texttt{\$t9} for the
callee address in dynamically linked calls. The canonical call sequence is:

\noindent\begin{minipage}{\linewidth}
\begin{lstlisting}[label={list:mipspic}]
la   $t9, strcmp
jalr $t9 ; strcmp
\end{lstlisting}
\end{minipage}

Because \texttt{jalr~\$t9} is the \emph{standard} call idiom on MIPS rather than an exceptional indirect dispatch, naively treating it as an unknown indirect call would flood the runtime discovery mechanism (Section~\ref{sec:impl:indirect}) at nearly every function call, creating significant overhead. \ourtool{}'s IDA script therefore recognises the \texttt{\$t9}-load / \texttt{jalr~\$t9} pattern in both its forms: an immediate symbol load (\texttt{la~\$t9,~sym})
and a pointer load from the Global Offset Table (GOT) or a data slot (\texttt{lw~\$t9,~\ldots}) whose cross-reference resolves to a named target. Either form is recorded as a statically known call. This leaves Section~\ref{sec:impl:indirect}'s dynamic mechanism to handle only dispatch sites whose callee cannot be resolved statically, such as function-pointer tables populated at runtime.

\paragraph{Distance score computation.}
\label{sec:impl:distance}%
Given the CFGs, CG, and call-site map, \ourtool{} computes per-block distances using the same shortest-path formulation as \aflgo{}~\cite{dgf1}: for each basic block $b$, the score is the harmonic mean of the shortest-path distances from $b$ to each target block, computed over the linked CFG-plus-CG graph. The DBB classifier (Section~\ref{sec:dbb-classifier}) is applied as part of this pass, and each output record includes a flag indicating whether the block is a DBB.

\aflgo{}'s distance script resolves a block identifier to its graph node by scanning all CFG nodes linearly per query, yielding $O(|V|^2)$ total cost for $|V|$ blocks. \ourtool{} replaces this with a pre-built hash index: at startup, the range of each block is inserted into a hash table keyed by each component address.  Subsequent lookups are $O(1)$, removing the quadratic term for firmware binaries with tens of thousands of blocks.

\subsection{QEMU-Based Runtime Instrumentation}
\label{sec:impl:qemu}

Distance-annotated basic blocks must be tracked at runtime to compute per-seed distances. Since recompilation is unavailable, \ourtool{} extends QEMU's TCG translation layer.

\paragraph{TCG hook for distance tracking.}
AFL's QEMU from \firmafl{} already instruments every Translation Block (TB) translation to invoke
\texttt{afl\_maybe\_log}, which updates the shared edge-coverage bitmap. \ourtool{} extends this
hook: when the TB's address matches an entry in the preloaded distance map, it increments a per-DBB hit counter and, at the end of each execution, combines the per-DBB distances into a single per-seed distance: each DBB's static distance is first adjusted by its satisfiability score ($d_i^\star = d_i / (1 + 99\,\sigma_i)$), and the adjusted values are combined by reciprocal-sum, as described in Section~\ref{sec:distance}.
This instrumentation is confined to the user-mode QEMU instance, where the binary's basic blocks are actually translated and executed. \firmafl{}'s system-mode instance serves only system calls and page-fault resolution against the full-system state; it never re-translates the binary's control flow, so annotating it would record kernel blocks that carry no proximity-table entry and add overhead without signal.

\paragraph{Branch-direction instrumentation.}
A second TCG hook is inserted during conditional-branch translation.  When QEMU processes a conditional branch, its TCG IR contains a \texttt{bcond} variable holding the boolean outcome of the condition. The hook fires only when the branching block is itself a classified DBB. It then compares the distances of the two successor addresses: the closer one identifies the path toward the target, and the required branch direction for DBB~$i$ and the direction actually taken are each written into dedicated per-DBB slots in shared memory. These values are read by the fuzzer after each execution to update satisfiability scores via the algorithm described in
Section~\ref{sec:sat-score}.

\paragraph{Shared-memory layout.}
The shared-memory region extends the standard AFL bitmap with two additional segments: a distance-accumulator block (distance sum and hit count over the proximity-annotated blocks), and a hit-count block with one saturating byte per distance-table entry, bumped whenever the corresponding DBB or target block executes. The branch directions live in two further segments, indexed the same way. The trim check (see Section~\ref{sec:impl:trim}) hashes the entire extended region to preserve DBB progress across seed minimisation.

\paragraph{Indirect call discovery.}
\label{sec:impl:indirect}%
Static analysis cannot resolve indirect calls whose targets depend on runtime state. \ourtool{} addresses this gap via a hybrid strategy.

The \emph{static component} handles the dominant case on MIPS: the PIC convention pattern
described in Section~\ref{sec:impl:extraction} is detected during the disassembler pass and
incorporated into the statically known call graph.

The \emph{dynamic component} handles the remaining indirect transitions: the translator tags each TB ending in an indirect call or computed jump, and \texttt{afl\_maybe\_log} carries the tag to the next block, pairing the tagged predecessor with
its successor. Each pair resolves to its enclosing functions and is kept unless intra-procedural, the reverse of a known edge, or already in the preloaded edge set. The new edge is recorded in a hash map and appended to a file in the working directory.

Because new edges may alter shortest-path distances to the target, an external watcher process polls, at configurable intervals, a flag the fuzzer raises once a logged edge could actually shorten some function's distance to a target; the other edges are still recorded and are folded into the next pass. The watcher then reruns the distance-computation pass and signals completion; the fuzzer loads the new table into the inactive half of a double-buffered region and swaps generations at the next queue-entry boundary, without pausing.

\subsection{Distance-Aware Queue Ordering}
\label{sec:impl:queue}

Standard coverage-guided greybox fuzzing stores seeds in a first-in-first-out queue, which is acceptable when all seeds contribute equally to coverage. In a directed campaign, seeds close to the target should be prioritised: each fuzzing cycle should begin from the most promising seed available.

\ourtool{} maintains seeds in a sorted queue ordered by ascending $D(q)/w_i$, where $w_i$ is the weight of the seed's most under-served target~$i$ (Section~\ref{sec:stagnation}), so that lower distance and a more
under-served target both mean higher priority. To prevent starvation of far-from-target seeds, each cycle re-sorts the queue and marks every entry unhandled, then fuzzes each entry exactly once regardless of its rank, while a newly inserted seed that outranks the current one preempts it. This hybrid policy maintains rapid progress on promising paths while ensuring all seeds eventually receive mutations.

\paragraph{Per-target reach tracking.}
The saturation and fairness mechanisms of Section~\ref{sec:stagnation} are driven by three per-target counters maintained on the fuzzer side: the cumulative number of executions reaching each target, a per-cycle count of scheduling visits, and the wall-clock timestamp of the most recent saved input reaching it. They are consulted by the energy-weighting routine (\texttt{saturation\_factor})
and the starvation-aware queue ordering. A per-target, per-cycle mutation cap bounds the energy any single target can absorb within one cycle, ensuring the schedule revisits neglected targets.

\subsection{Mutation Optimisations}
\label{sec:mut-opt}

\paragraph{Satisfiability-guided splicing.}
AFL's recombination operator selects a splice partner uniformly at random from seeds whose content
differs from the current seed.  \ourtool{} replaces this with a targeted strategy: for the first
DBB~$i$ at which the current seed has a zero satisfiability score, the splice partner is selected as
the corpus seed with the highest satisfiability score at that specific DBB. The hybrid keeps the current seed's overall structure while importing the bytes that drive the predicate it is stuck on. When no failed DBB has a distinct leader long enough to splice, the operator falls back to AFL's random choice.

\paragraph{Satisfiability-Preserving Mask (SPM): preventing mutation regression.}
\label{sec:sniff-mask}%
A characteristic failure mode of directed fuzzing occurs when mutation destroys a comparison
constraint that a seed had previously satisfied. Suppose CmpLog-guided mutation has found that the cookie key must be exactly \texttt{uid}, producing a seed with
\texttt{HTTP\_COOKIE=uid=<payload>} and satisfiability score~1.0 at the corresponding DBB. In the subsequent havoc stage, any random mutation touching the bytes covering \texttt{uid=} may destroy this solved constraint, forcing the fuzzer to re-solve the comparison from scratch. Re-solving is expensive: it requires CmpLog-guided mutations to re-discover the correct string, competing against random mutations that may revert the solution at each round.

\windranger{} implements a superficially similar byte-sensitivity analysis, its \emph{cb\_mask}. Its semantics are exploratory: the mask identifies bytes whose mutation \emph{triggers} a DBB and
biases mutation \emph{toward} them, concentrating effort on positions that are relevant to
reaching new DBBs. \ourtool{}'s mask is preservative: it identifies bytes whose mutation
\emph{destroys} an already-satisfied condition and biases mutation \emph{away} from them.  The two mechanisms are complementary.

The SPM is computed once per seed at the start of the mutations routine via a lightweight single-flip sensitivity analysis, skipped for seeds above a length bound or reaching no DBB. The seed is executed once to establish a
baseline, recording for each DBB~$i$ whether it was hit and whether the correct branch direction was taken. Then, for each input byte~$j$, byte~$j$ is flipped by XOR~0xFF, the modified input is executed, and byte~$j$ is marked protected if any of the following holds:

\begin{enumerate}[label=(\alph*)]
  \item A DBB that was hit in the baseline is no longer hit (the byte controls a structural entry condition to the DBB);
  \item A condition that was satisfied in the baseline is no longer satisfied (the byte directly affects the comparison operand).
\end{enumerate}

Case~(a) captures structural bytes such as the HTTP method name or the cookie key prefix; case~(b)
captures semantic bytes such as the value portion of a known comparison, which CmpLog-guided
mutation may have already placed correctly. The total cost of the mask pass is $|q| + 2$ executions: one baseline, $|q|$ flip-and-restore passes, one restore run.

If more than half of the input bytes are marked protected, the mask is discarded to prevent the seed from being frozen. Protected positions are skipped in all deterministic mutation stages and excluded from the pool of candidate positions in havoc.

\paragraph{Input-aware mutation.}
\label{sec:pattern-mutators}%
\ourtool{}'s HTTP socket and CGI environment-variable channels both carry structured input, subject to the constraints described in Section~\ref{sec:back:grammar}: RFC~7230 request syntax and the POSIX key-value convention~\cite{posix_env} respectively. Unrestricted byte-level mutation frequently destroys these structures, causing inputs to be rejected before reaching any meaningful processing logic.

For HTTP inputs, \ourtool{} integrates the Grammar Mutator~\cite{grammar_mutators}, an AFL++ custom mutator that enforces adherence to user-defined grammars while retaining the
feedback-driven mutation loop.  Grammar-aware mutations produce structurally valid HTTP packets
that exercise the payload-processing logic rather than being discarded by the request parser.

For environment variable inputs, which follow a key-value format, \ourtool{} takes a different approach, since the Grammar Mutator's overhead and expressiveness are disproportionate to the limited structural complexity of these inputs. It leverages a lightweight LLM-assisted grammar extraction pipeline: a locally hosted Qwen2.5-14B model~\cite{qwen25} (via Ollama~\cite{llm2}) is given a curated list of uppercase, digit, and underscore strings extracted from the binary under test, which form candidate environment variable keys following POSIX conventions. The LLM is then asked, through a two-stage few-shot prompting workflow, to identify which strings are likely to represent environment variable keys. The identified keys are then used to produce PCRE2 regexs of the form \texttt{KEY=VALUE}, capturing the mutable region (the value) and the
immutable region (the key prefix), to allow the fuzzer to confine mutations to the value while preserving the structural key.

Each havoc stage draws an explore/exploit switch whose exploit probability grows as the seed's distance shrinks, reaching certainty for seeds that already reach a target: in exploit mode, mutations are restricted to the value region identified by the regex; in explore mode both are waived, allowing unrestricted byte-level mutation to exercise the key-parsing logic. This switch ensures that structure-breaking mutations are not eliminated entirely, since malformed key-value pairs may exercise error-handling paths that valid inputs bypass.

The satisfiability-preserving mask interacts directly with this mechanism, which we refer to as \textsc{Pattern Mutator}: the effective mutable position pool is the intersection of the pattern range (from the regex) and
the safe positions (from the SPM). A position must be both inside the mutable
region and free of a solved constraint to be eligible for mutation. In contrast, the Grammar Mutator regenerates the buffer, so positions cannot be preserved; to address this, \ourtool{} extracts the protected byte runs from the parent and re-inserts any that the generated input has dropped, anchored on the surrounding content.

\subsection{AFL Enhancements}
\label{sec:impl:afl-correctness}

Two aspects of AFL's standard infrastructure require adaptation to preserve the accuracy of
\ourtool{}'s distance and satisfiability tracking.

\paragraph{Extended trim hash.}
\label{sec:impl:trim}%
AFL's seed minimisation (trim) phase reduces seed size by removing bytes that do not change the coverage bitmap hash. The standard hash covers only the first \texttt{MAP\_SIZE} bytes of the shared-memory region, which records edge coverage. In \ourtool{}, a trim that removes bytes critical to reaching a DBB might not change the edge bitmap (if those bytes lie on a
previously seen edge) yet would destroy the DBB-hit flags and condition bits that underpin
satisfiability scoring. The result would be a trimmed seed that appears to have a high satisfiability score at
DBBs it can no longer reach. 

\ourtool{} extends the trim hash to cover the whole instrumentation region, including the distance-accumulator block and the DBB-flag block. A byte removal is accepted only if the hash of that entire region is unchanged, so no DBB that the seed reached can be silently lost to minimisation.
When a trim is accepted, the satisfiability-preseving mask (Section~\ref{sec:sniff-mask}) is invalidated because byte removal shifts the positional interpretation of all subsequent mask entries. The mask is recomputed at the start of the next fuzzing cycle on the trimmed seed, at the cost of $|q| + 2$ additional executions.

\paragraph{Hash collision mitigation.}
\label{sec:impl:hash}%
AFL's standard coverage-bitmap hashing scheme compresses a TB address \texttt{addr} into a
bitmap index via:
\[
  \texttt{loc} = \bigl((\texttt{addr} \gg 4) \oplus (\texttt{addr} \ll 8)\bigr)
    \mathbin{\&} (\texttt{MAP\_SIZE} - 1).
\]
This scheme is optimised for speed but not collision resistance: multiple distinct TBs may map
to the same bitmap slot. For edge coverage, collisions are acceptable approximations; for
satisfiability score tracking, a collision causes the satisfiability score of an unrelated block to be
incorrectly attributed to a DBB, distorting prioritisation.

\ourtool{} replaces the standard hashing scheme for every address-to-slot lookup with xxHash~\cite{xxhash}, a non-cryptographic hash function with near-uniform distribution and very high throughput. The performance overhead relative to the bit-shift scheme is negligible compared to the per-execution cost of CmpLog operand capture, and the change eliminates the systematic attribution errors that degrade satisfiability accuracy.

\subsection{Bug Detection}
\label{sec:bug-oracles}

AFL's standard crash oracle relies on the guest process's own fault mechanism: a crash is recorded only when the process dies on a signal. This creates a visibility gap for \emph{silent} overflows, where the corrupted bytes land in an already-mapped writable region and the process continues without triggering a fault. A write past a fixed-size global, such as a \texttt{strcpy(\&timestamp,~attacker\_input)} overflowing into adjacent \texttt{.bss} or \texttt{.data} bytes, produces no observable signal: the request handler finishes normally and the forkserver records a clean exit.  \ourtool{} closes this gap with two complementary oracles.

\paragraph{Global-variable overflow oracle.}
The oracle intercepts a set of libc functions whose write is unbounded or whose bound is
attacker-controlled: \texttt{strcpy}, \texttt{strcat}, \texttt{sprintf}, \texttt{gets},
\texttt{strncpy}, \texttt{strncat}, \texttt{vsprintf}, \texttt{memcpy}, and \texttt{memmove}. At function entry, the hook reads
\texttt{dst} and \texttt{src} from the calling-convention argument registers, computes the
effective write length per call family (\texttt{strlen(src)} for unbounded copies,
\texttt{strlen(dst)+strlen(src)} for \texttt{strcat}, the explicit \texttt{n} for bounded variants), and signals a bug if that length exceeds the destination capacity.

Capacity is resolved through three mechanisms, with the exact-size checks taking precedence. At fork-server startup, \ourtool{} walks the binary's \texttt{.dynamic}/\texttt{.dynsym}
tables (mapped directly into host memory inside the user-mode emulation layer) and records
every object-type symbol as an \texttt{(addr,~size)} pair. When \texttt{dst} resolves to a
known object, the overflow condition is $\mathit{needed} > \mathit{obj.end} - \mathit{dst}$,
where $\mathit{needed} = \mathtt{strlen}(\mathit{src})+1$ for unbounded families. When \texttt{dst} instead falls inside a tracked heap chunk, the chunk's exact remaining size is
used. Static globals with neither a \texttt{.dynsym} entry nor a tracked chunk fall back to a configurable length threshold $\mathit{needed} > N$, applied only to the unbounded families, since for bounded copies a large explicit length argument is normal and flagging it yields false positives. A twin pass over the C library's own dynamic symbols resolves the function entry points used to intercept calls at runtime. When an overflow is detected, the child exits with a sentinel exit code, which AFL, running the binary under QEMU, classifies directly as a crash.

This mechanism follows the same structure as AddressSanitizer~\cite{asan} libc interceptors, which wrap each string function and check the access range before invoking the real implementation. The key difference is scope: \ourtool{}'s per-object bound is
\texttt{.dynsym}-based, giving precise coverage for externally linked globals but narrower
total coverage than ASan's shadow-memory redzones.

\paragraph{Heap overflow oracle.}
Heap overflows share the same fault-visibility gap: \texttt{recv(fd,~buf,~attacker\_len,~0)} overflowing into the next chunk also lands inside an already-mapped writable region. There
is an additional timing problem: the corruption is only detected when the next
\texttt{malloc}/\texttt{free} reads the poisoned chunk header, which typically occurs in a
later execution after the forkserver has moved on to the next input.

\ourtool{} uses three detection strategies:

\begin{enumerate}

  \item \emph{brk-anchored chunk walk.} Walking the chunk chain (masking the three flag bits,
  advancing by \texttt{size}) and checking for implausible values ($\mathit{size} < 8$,
  $\mathit{size} > 256$~MB, or $\mathit{ptr}+\mathit{size} > \mathit{brk}$) detects corruption that overwrites the next chunk's size field. The arena bounds
  (\texttt{mm->start\_brk}, \texttt{mm->brk}) are kernel-side fields; \ourtool{} derives their offsets from \texttt{task\_struct} at snapshot time and emits a \texttt{HEAP:start:brk}
  record that the walker consumes. The walk runs after each I/O syscall that can deliver attacker
  data (\texttt{read}, \texttt{write}, \texttt{recv}, \texttt{recvfrom}) and at normal process exit, adding approximately 2--5~$\mu$s per invocation with no measurable AFL
  throughput loss. To guard against a format mismatch, the walker is calibrated once against a known-clean heap at startup (calibration result is tri-state: \emph{untrusted} /
  \emph{disabled} / \emph{trusted}): a failed calibration disables the oracle rather than producing spurious crashes.

  \item \emph{malloc/free redzone.} Using the same \texttt{.dynsym} resolver extended to
  \texttt{malloc}/\texttt{free}/\texttt{calloc}/\texttt{realloc}, \ourtool{} bumps every
  requested allocation size by 16~bytes, stamps the extra bytes with the sentinel value
  \texttt{0xA5}, and verifies the trailer at each \texttt{free} or \texttt{realloc}, as well as
  sweeping the trailers of all live chunks at normal process exit. This strategy catches what the first strategy structurally cannot: overflows of the \emph{last} chunk in the
  arena (no successor header to corrupt) and short overflows that land in \texttt{dlmalloc}'s
  internal padding rather than the next header.

  \item \emph{copy-time chunk bound.}  When the destination of an intercepted copy falls inside a tracked chunk, the effective write length is compared against that chunk's exact remaining size at the call site. This is the only strategy that reports a chunk that is smashed but never freed and whose corruption never reaches a successor header, a case that otherwise leaves the execution recorded as clean.

\end{enumerate}

None of these strategies detects use-after-free or double-free, and
\texttt{mmap}-backed allocations larger than \texttt{M\_MMAP\_THRESHOLD} bypass the brk-arena walker entirely.

\paragraph{Length-extension mutation.}
\label{sec:len-extend}%
A buffer overflow triggered by an attacker-controlled copy requires the fuzzer to produce inputs that are both structurally valid \emph{and} long enough.  Coverage-guided fuzzers do not naturally grow inputs toward an overflow: intermediate seeds that reach the target basic block with a near-overflowing copy length but no new coverage edge are discarded, so the
queue never accumulates the length gradient needed to trigger the bug.

\ourtool{} addresses this through \emph{length-extension mutation}, which reuses the overflow oracle's instrumentation. The libc-copy hooks already compute the \texttt{(copy\_len, capacity)} pair at the point and time where it is most meaningful. This pair has the same shape as a CmpLog comparison entry: it expresses a
quantitative relationship between an attacker-influenced value (\texttt{copy\_len}) and a
fixed bound (\texttt{capacity}).  \ourtool{} injects it as a \emph{synthetic} CmpLog record tagged with a reserved sentinel byte so that the consumer can distinguish it from a real instruction comparison.

The AFL-side consumer extends RedQueen's colorization stage with a branch that recognises this
tag and, instead of substituting bytes to satisfy a value equality, \emph{clones} the colorization taint region covering \texttt{src} enough times to push \texttt{copy\_len} past
\texttt{capacity}. Concretely, given $\mathit{copy\_len} = L$, $\mathit{capacity} = C$, and a cloned taint region of length $R$, the consumer computes $\mathit{copies} = \lceil (C - L + 9) / R \rceil$ and inserts $\mathit{copies}$ repetitions of that region immediately after it, before re-executing the grown input. If the grown input now overflows the destination, the overflow oracle fires on the same execution.

Two limitations apply. First, the technique requires that the relevant input bytes appear in
the colorization taint set; if \texttt{src}'s bytes were never traced, the extension probe
is a no-op. Second, the synthetic comparison is 32-bit only, which is sufficient for all targets in the evaluation corpus.


\section{Evaluation}
\label{sec:evaluation}

We evaluate \ourtool{} on a benchmark of 40 firmware vulnerability sites drawn from real-world IoT devices. Our evaluation is organised around five research questions:

\begin{itemize}

    \item \textbf{RQ1 (Effectiveness).} Does \ourtool{}, operating purely on stripped vendor binaries (\ref{p1}), reproduce firmware vulnerabilities faster than state-of-the-art coverage-guided and directed greybox fuzzers, and than a state-of-the-art firmware re-hosting framework?

    \item \textbf{RQ2 (Directedness).} Do \ourtool{}'s semantic refinements, i.e., the three-condition DBB classification (\ref{p2}) and the satisfiability-adjusted distance metric (\ref{p3}), improve target-reaching effectiveness over the other baseline directed greybox fuzzers?

    \item \textbf{RQ3 (Energy allocation).} Does \ourtool{}'s power scheduling (\ref{p4}) allocate energy better than a single global annealing clock, spatially, across co-located targets, and temporally, across DBBs discovered late in a campaign?

    \item \textbf{RQ4 (Component contribution).} How much does each component of \ourtool{} contribute to its overall performance, and does removing it cost what the design predicts?

    \item \textbf{RQ5 (Case Studies).} Three case studies examine up close how \ourtool{}'s mechanisms behave on individual targets.

\end{itemize}

The remainder of this section describes the experimental setup (Section~\ref{sec:eval:setup}) and then addresses RQ1--RQ5 in turn, closing each with an explicit answer.

\subsection{Experimental Setup}
\label{sec:eval:setup}

\paragraph{Infrastructure.}
All experiments run on a dual-socket Intel Xeon E5-4610~v2 server (16 physical cores / 32 hardware threads at 2.30~GHz) with 256~GiB of RAM, under Ubuntu~24.04.1 LTS. Following common practice for fair fuzzer comparison~\cite{klees2018evaluating}, each fuzzing campaign runs in its own privileged Docker container pinned to a single physical core (both hardware threads, via \texttt{cpuset}) with \texttt{AFL\_NO\_AFFINITY} set, and up to 15 campaigns execute in parallel, leaving one core for orchestration and the host.

\begin{table}[!tp]\centering\scriptsize
  \setlength{\tabcolsep}{3pt}
  \renewcommand{\arraystretch}{1.15}
  \caption{Firmware images in the evaluation dataset.}
  \label{tab:firmwares}
  \begin{adjustbox}{width=\linewidth}
  \csvreader[separator=semicolon, head to column names, head to column names prefix=col,
    tabular=|l|l|l|c|c|,
    table head=\hline \textsc{ID} & \textsc{Vendor} & \textsc{Model} & \textsc{Version} & \textsc{Arch}\\\hline\hline,
    late after line=\\\hline,
    table foot=]%
    {data/DatasetFirmware.csv}{}%
    {\colid & \colvendor & \colmodel & \colversion & \colarch}
  \end{adjustbox}
\end{table}

\paragraph{Dataset.}
We assemble a benchmark of 32 closed-source firmware images drawn from four vendors (D-Link, Netgear, TRENDnet, and TP-Link), listed in Table~\ref{tab:firmwares}, and emulated through Firmadyne~\cite{firmadyne}, spanning the little- and big-endian MIPS architectures (\textsc{mipsel} and \textsc{mipseb} respectively). From these images we derive 40 directed targets, each paired with the virtual address of a vulnerable sink and reached over one of three input channels: a TCP socket for HTTP services, an environment variable for CGI handlers, and a file stream for configuration parsers. \firmafl{} exposes only the first two; \ourtool{} adds three more: the file stream above, a UDP datagram, and the buffer returned by an \texttt{ioctl} on a device node. Each target address is obtained by manually locating the vulnerable sink in the binary disassembled with IDA~\cite{ida}, guided, where available, by the CVE advisory or a public proof-of-concept, and recording its relative load-base offset for the static pipeline (Section~\ref{sec:impl:extraction}); this step is one-off and manual, consistent with the reachability assumptions discussed in
Section~\ref{sec:limitations}.

The choice of \emph{which} basic block to designate as the target is itself consequential~\cite{weissberg2024targetselection}:
a target on a hot path shared by most inputs carries little discriminative signal, since nearly every seed reaches it and the distance metric can no longer
separate promising inputs from the rest. We therefore designate as the target the most discriminative basic block that every triggering input must traverse. Where the vulnerable sink lies on a cold, guarded path this is the sink itself; where the sink sits on a hot path we instead select an upstream block, typically the condition gating entry to the vulnerable control flow (a header-name or request-path check), that isolates the input characteristic required to reach the sink, so that the seeds prioritised toward the target are exactly those still able to drive the corruption downstream.

Table~\ref{tab:cves} enumerates the targets. Of the 40, 26 carry an assigned CVE and two an EDB-ID; the remaining 12 have no public identifier at the time of writing. Of these, five were reproduced from public crash reports (labelled \textsf{housefuzz-1} to \textsf{housefuzz-5}) and seven newly identified in this work (labelled \textsf{bullseye-1} to \textsf{bullseye-7}); we withhold all proof-of-concept inputs until each has been disclosed to the affected vendor, or to \textsc{CERT/CC} where the device is past end of life.

\begin{table}[!tp]\centering\scriptsize
    \setlength{\tabcolsep}{4pt}
    \renewcommand{\arraystretch}{1.15}
    \caption{The 40 directed targets. IDs refer to Table~\ref{tab:firmwares}.}
    \label{tab:cves}
    \begin{adjustbox}{width=\linewidth}
    \csvreader[separator=semicolon, head to column names, head to column names prefix=col,
      tabular=|l|c|l|c|l|,
      table head=\hline \textsc{Reference} & \textsc{ID} & \textsc{Binary} & \textsc{Channel} & \textsc{Weakness}\\\hline\hline,
      late after line=\\\hline,
      table foot=]%
      {data/DatasetCVE.csv}{}%
      {\colref & \colid & \colbinary & \colchannel & \colimpact}
    \end{adjustbox}
\end{table}

\paragraph{Seed corpus.}
The initial seed corpus for all targets is produced with a local LLM, Qwen2.5-14B~\cite{qwen25}, served through Ollama: valid HTTP packets for socket targets, extracted environment-variable keys for CGI targets, and a single valid configuration file for the file-stream target. The same corpus is shared across all six configurations, so that any performance difference stems from the fuzzing strategy rather than the starting inputs.

\paragraph{Configurations.}
We compare \ourtool{} against five baselines at two levels. Four are controlled baselines that span the coverage-guided and directed greybox fuzzing spectrum and share \ourtool{}'s substrate: they run on top of the same \firmafl{}~\cite{firmafl} dual-QEMU emulation back-end and differ only in their fuzzing strategy, isolating the effect of the directed-fuzzing logic. \firmafl{} is the unmodified back-end, i.e., stock coverage-guided fuzzing. \firmafl{}\textsuperscript{\textdagger} additionally enables CmpLog comparison-guided mutation~\cite{cmplog} and grammar-aware mutation, but remains undirected. \textsc{Firm-AFL-Go} implements \aflgo{}'s aggregate structural distance~\cite{dgf1}, while \textsc{Firm-AFL-WR} implements \windranger{}'s deviation-basic-block guidance~\cite{wind}; both build on the \firmafl{}\textsuperscript{\textdagger} mutation stack so that any difference relative to \ourtool{} stems from directed fuzzing heuristics rather than the mutator, with all directed configurations being driven by the same target specification. Both follow the published algorithms rather than the original implementations: \aflgo{} and \windranger{} derive their graphs from LLVM~IR at compile time, whereas here both distances are computed over control-flow graphs and a call graph recovered statically from the stripped binary and applied through the same QEMU instrumentation as \ourtool{}. \textsc{Firm-AFL-WR} selects DBBs by \windranger{}'s structural criterion (Section~\ref{sec:back:dbb}) and, lacking the source-level data-flow analysis behind \windranger{}'s $\Psi$ weight, approximates it from the comparison operands that change between a seed and its colorized twin. Since four of the five \firmafl{}-based configurations rely on CmpLog to resolve hard-coded comparisons automatically, we supply stock \firmafl{} with a keyword dictionary of string tokens, injected during AFL's dictionary mutation stage, so that it can solve the same comparisons: without it the baseline times out on most targets, yielding an uninformative comparison. This mirrors \firmafl{}'s own distribution, which ships 10 sample targets each accompanied by such a dictionary; we generate the equivalent files for the new dataset entries by extracting the human-readable strings from each binary under test. 

\textsc{Greenhouse}~\cite{jun2023greenhouse} is the fifth competitor in our evaluation, an end-to-end baseline rather than a controlled one: it re-hosts each firmware image with its own re-hosting pipeline rather than sharing \firmafl{}'s back-end, and fuzzes undirected. Its input channel is also narrower: \textsc{Greenhouse} drives a re-hosted service over a TCP socket only, so we attempt re-hosting solely for the 27 HTTP targets. Comparing on the environment-variable and file targets would have measured \ourtool{}'s channel-specific injection against a pipeline that cannot deliver those inputs at all, which would understate \textsc{Greenhouse} rather than inform the comparison. Of the 27 attempted, 18 re-host successfully; the other 22 of the 40 targets carry no \textsc{Greenhouse} entry in Table~\ref{tab:rq1} and are excluded from every \textsc{Greenhouse} comparison. Each (configuration, target) pair is run 10 times for 10~h, and the three directed configurations share \aflgo{}'s implementation default for the annealing cut-off, $t_x = 10$~minutes.

\paragraph{Metrics.}
The primary metric is the \emph{Time-to-Exposure} (TTE), measured as wall-clock time; a run that does not trigger the vulnerability within the 10-hour budget is assigned a TTE equal to the budget, following \aflgo{}~\cite{dgf1}. Such runs are censored observations whose true TTE is unknown, which makes the Mann--Whitney U test inapplicable~\cite{kim2024evaluating}; we therefore report effect sizes and speed-up factors rather than $p$-values. As is standard in directed greybox fuzzing work~\cite{dgf1,wind}, for each (configuration, target) pair we report the mean TTE ($\mu$TTE) over the 10 runs, the speed-up factor relative to each baseline, and the Vargha--Delaney $\hat{A}_{12}$ statistic. The speed-up factor is the ratio of the baseline $\mu$TTE to the configuration's $\mu$TTE, so that values above $1.0$ indicate a faster configuration. The $\hat{A}_{12}$ statistic is a non-parametric effect-size measure~\cite{vargha2000critique}, the recommended standard for comparing randomised algorithms~\cite{arcuri2014hitchhiker}: over the TTE values from the baseline and a given configuration, it estimates the probability that the configuration produces a smaller TTE than the baseline, so values above $0.5$ favour the configuration and values below it favour the baseline, with $0.5$ indicating no difference and $0.71$ the conventional threshold for a large effect.

\paragraph{Attribution and ablation logging.}
RQ2--RQ5 are evaluated on a 10-image subset of the benchmark (Table~\ref{tab:subset}): eight images drawn at random from the full benchmark and two (images~25 and~26) chosen deliberately as multi-target test cases, so that the components whose behaviour only manifests under multiple co-located targets are exercised. Two instrumented run sets are collected on it, three runs of 10~h per configuration, kept separate from the RQ1 campaign so they never perturb its timing: The first is \emph{attribution logging}: recording for every scheduling visit which stage produced each seed or crash, the scheduling state at that point, and the set of directed targets it lit. The second is \emph{ablation logging}: the six leave-one-out configurations of Section~\ref{sec:eval:rq4}, together with the full system, are run to record, per queued seed and crash, the elapsed time and execution count needed to reach and to trigger each target. Both instrumented sets pay for this detail in throughput: the logging sits on the execution path, so their absolute times are systematically longer than those of the RQ1 campaign and are not comparable with it.

\begin{table}[t]\centering\small
    \setlength{\tabcolsep}{4pt}
    \renewcommand{\arraystretch}{1.15}
    \caption{The 10-image RQ2–RQ5 subset. IDs refer to Table~\ref{tab:firmwares}.}
    \label{tab:subset}
    \csvreader[separator=semicolon, head to column names, head to column names prefix=col,
      tabular=|l|l|l|c|c|,
      table head=\hline \textsc{ID} & \textsc{Vendor} & \textsc{Model} & \textsc{Binary} & \textsc{Arch}\\\hline\hline,
      late after line=\\\hline,
      table foot=]%
      {data/DatasetSubset.csv}{}%
      {\colid & \colvendor & \colmodel & \colbinary & \colarch}
\end{table}

\subsection{RQ1: Effectiveness}
\label{sec:eval:rq1}

RQ1 asks whether \ourtool{}, working only on stripped vendor binaries (\ref{p1}), reproduces firmware vulnerabilities faster than state-of-the-art coverage-guided and directed greybox fuzzers, and than a state-of-the-art re-hosting framework. We run every (configuration, target) pair 10 times for 10~h and record the time-to-exposure (TTE). Table~\ref{tab:rq1} reports, per target, the mean TTE ($\mu$TTE) of \ourtool{}, of the four \firmafl{}-based baselines, and of \textsc{Greenhouse}~\cite{jun2023greenhouse}, together with \ourtool{}'s speed-up factor and Vargha--Delaney $\hat{A}_{12}$ against each of them; Table~\ref{tab:rq1_summary} summarises those comparisons per configuration: targets triggered, geometric-mean speed-up, and mean $\hat{A}_{12}$ ($\mu\hat{A}_{12}$).

\begin{table*}[!tp]\centering\scriptsize
    \caption{Per-target $\mu$TTE over 10 10-hour runs, with \ourtool{}'s speed-up factor and $\hat{A}_{12}$ against each configuration; bold marks the lowest $\mu$TTE per target.}
    \label{tab:rq1}
    \setlength{\tabcolsep}{4pt}
    \renewcommand{\arraystretch}{1.15}
    \resizebox{\textwidth}{!}{%
    \csvreader[separator=semicolon, head to column names, head to column names prefix=col,
      tabular=|l|ccc|ccc|ccc|ccc|ccc|c|,
      table head=\hline
        \textsc{Target} & \multicolumn{3}{c|}{\textsc{Firm-AFL}}
          & \multicolumn{3}{c|}{\textsc{Firm-AFL}\textsuperscript{\textdagger}}
          & \multicolumn{3}{c|}{\textsc{Firm-AFL-Go}}
          & \multicolumn{3}{c|}{\textsc{Firm-AFL-WR}}
          & \multicolumn{3}{c|}{\textsc{Greenhouse}}
          & \multicolumn{1}{c|}{\ourtool{}}\\\cline{2-17}
        & $\mu$TTE & \textsc{Fac.} & $\hat{A}_{12}$
          & $\mu$TTE & \textsc{Fac.} & $\hat{A}_{12}$
          & $\mu$TTE & \textsc{Fac.} & $\hat{A}_{12}$
          & $\mu$TTE & \textsc{Fac.} & $\hat{A}_{12}$
          & $\mu$TTE & \textsc{Fac.} & $\hat{A}_{12}$
          & $\mu$TTE\\\hline\hline,
      late after line=\\\hline,
      table foot=]%
      {data/RQ1_results.csv}{}%
      {\colref
        & \colvanmu & \colvanf & \colvana
        & \colplusmu & \colplusf & \colplusa
        & \colgomu & \colgof & \colgoa
        & \colwrmu & \colwrf & \colwra
        & \colghmu & \colghf & \colgha
        & \colbullmu}%
    }%
\end{table*}

\begin{table}[!tp]\centering\small
      \setlength{\tabcolsep}{6pt}
      \renewcommand{\arraystretch}{1.2}
      \caption{Targets triggered by each baseline, with \ourtool{}'s speed-up and $\mu\hat{A}_{12}$ over it.}
      \label{tab:rq1_summary}
      \resizebox{\columnwidth}{!}{%
      \csvreader[separator=semicolon, head to column names, head to column names prefix=col,
        tabular=|l|c|c|c|,
        table head=\hline
          \textsc{Configuration} & \textsc{Triggered} & \textsc{Speed-up}
            & $\mu\hat{A}_{12}$\\\hline\hline,
        late after line=\\\hline,
        table foot=]%
        {data/RQ1_summary.csv}{}%
        {\colconf & \coltrig & \colfac & \colaeff}%
      }%
\end{table}

\paragraph{Reproduction reach.}
The headline result is coverage of the target set. \ourtool{} triggers \emph{all 40} targets within budget, against 35 for \textsc{Firm-AFL-WR}, 32 for \textsc{Firm-AFL-Go}, 29 for
\firmafl{}\textsuperscript{\textdagger}, and 17 for stock \firmafl{}. Most telling, five targets spanning three vulnerabilities (CVE-2019-11418, CVE-2018-19242, and CVE-2022-24355) are reproduced by \emph{no} \firmafl{}-based baseline, coverage-guided and directed alike: all four exhaust the full 10~h on them across all 10 runs of this campaign.

\paragraph{Time to exposure.}
Where a target is reached at all, \ourtool{} is typically fastest by a wide margin: its median $\mu$TTE over the 40 targets is under 10 minutes, an order of magnitude or more below every \firmafl{}-based baseline, whose medians run from under two hours to the full budget. Aggregated per baseline (Table~\ref{tab:rq1_summary}), the geometric-mean speed-up ranges from
$9.5\times$ against \textsc{Firm-AFL-Go} to $72.5\times$ against stock \firmafl{}, and the $\mu\hat{A}_{12}$ never falls below $0.81$, well above the $0.71$ large-effect threshold;
\ourtool{} carries a favourable effect size ($\hat{A}_{12}>0.5$) on at least 37 of 40 targets against every \firmafl{}-based baseline. Since non-triggering runs are charged the full budget rather than discarded, every factor is a lower bound.

\paragraph{Where directedness does not pay.}
Directedness recovers the search effort a target's guarding conditions would otherwise waste, so where a target has no such conditions there is nothing to recover and the distance and
scheduling bookkeeping is pure overhead. CVE-2018-10996 is the clearest case, a \texttt{session.cgi} environment-variable sink that almost any input drives, where \ourtool{} averages 74~s against 38~s for \textsc{Firm-AFL-Go}. Over the 40 targets a \firmafl{}-based baseline holds the lowest mean TTE on three, never by more than a few minutes; the fourth target on which \ourtool{} does not lead is CVE-2019-11418 on ID~14, lost to \textsc{Greenhouse} by 25~s, a re-hosting effect discussed below rather than a directedness one. \ourtool{}'s advantage concentrates where it matters: on targets whose guarding conditions make undirected reproduction improbable within the budget.

\paragraph{The \textsc{Greenhouse} column.}
\textsc{Greenhouse} is not an isolated-variable comparison, and its column moves for a different reason than the rest of the table. Because it re-hosts each image to run directly on the host rather than through \firmafl{}'s dual-QEMU scheme, it executes more test cases per second than any configuration sharing our back-end, its authors reporting a throughput 200\% higher than \textsc{EquAFL}~\cite{equafl}, itself a \firmafl{} successor. That throughput substitutes for guidance: it solves 16 of the 18 targets its pipeline supports, and it triggers three of the five targets no \firmafl{}-based baseline reaches at all. \ourtool{} is still faster on 17 of those 18, by a geometric mean of $9.8\times$, though by under $3\times$ on six of them, and it loses CVE-2019-11418 on ID~14 by $25$\,s. The comparison therefore measures two things at once, a directedness advantage and an emulation-throughput disadvantage, and we read it as an end-to-end result rather than as evidence about directedness.

\begin{answer}{1}
\ourtool{} reproduces \emph{all 40} directed targets within budget, against 35 for the strongest \firmafl{}-based baseline and 17 for stock \firmafl{}, with five targets reached by no \firmafl{}-based baseline. It is $9.5$ to $72.5\times$ faster by geometric mean than every such baseline, at a Vargha--Delaney effect size of $0.81$ or above. It also outpaces the end-to-end baseline \textsc{Greenhouse} on 17 of its 18 supported targets.
\end{answer}

\subsection{RQ2: Directedness}
\label{sec:eval:rq2}

\begin{figure*}[pos=htbp]    
    \centering \includegraphics[width=0.8\linewidth]{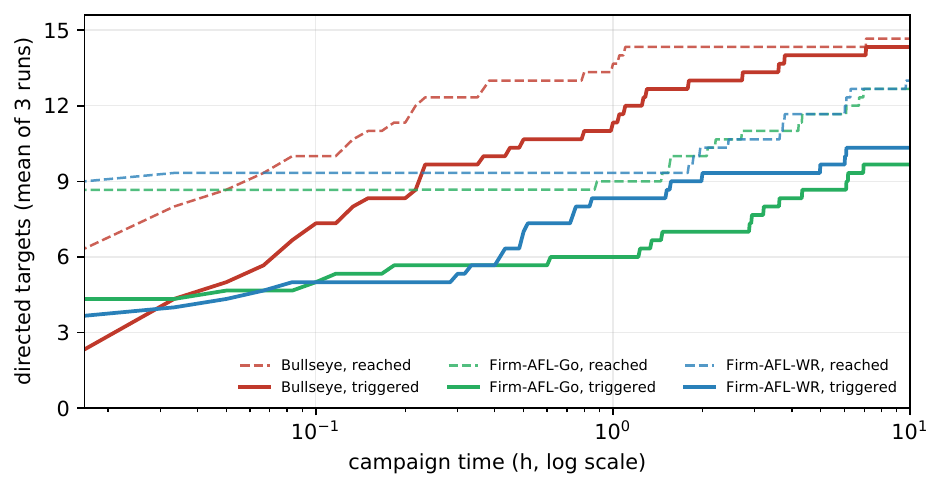}
    \caption{Directed targets reached and triggered over 10~h, averaged over three runs, across the instrumented subset.}
    \label{fig:rq2_targets}
\end{figure*}

RQ2 asks whether \ourtool{}'s semantic refinements, the three-condition DBB classification (\ref{p2}) and the satisfiability-adjusted distance (\ref{p3}), improve target-reaching effectiveness over \windranger{}'s structural DBBs and \aflgo{}'s aggregate distance. We answer it on the instrumented subset of Table~\ref{tab:subset}: 10 images carrying 15 directed targets (three of the images are multi-target), each fuzzed three times for 10~h per configuration with the per-visit event log enabled.

\paragraph{Target completeness.}
On this subset \ourtool{} reaches and triggers all 15 targets while \textsc{Firm-AFL-WR} triggers 12 and \textsc{Firm-AFL-Go} 10; Figure~\ref{fig:rq2_targets} traces run-averaged reach and trigger counts over campaign time: \ourtool{} starts behind both baselines, which reach the loosely guarded targets first and pay nothing for direction there (Section~\ref{sec:eval:rq1}), and
it is ahead of them from three minutes onward. Two of those targets, CVE-2018-19242 on ID~15 and CVE-2019-11418 on ID~31, are triggered \emph{only} by \ourtool{}: both baselines reach the target basic block in every single run and never convert that reach into a crash.

\begin{table}[!tp]\centering\small \setlength{\tabcolsep}{5pt} \renewcommand{\arraystretch}{1.2} 
    \caption{Deviation blocks \ourtool{}'s (BE) classifier flags beyond \textsc{Firm-AFL-WR}'s (WR), how many of them a run exercises, and the share of seed pairs each distance leaves tied.}
    \label{tab:rq2_p2}
    \resizebox{\columnwidth}{!}{%
    \csvreader[separator=semicolon, head to column names, head to column names prefix=col,
        tabular=|l|r|cc|cc|,
        table head=\hline
          \textsc{ID} & \textsc{Seeds} & \multicolumn{2}{c|}{\textsc{Added DBBs}}
            & \multicolumn{2}{c|}{\textsc{Tied seed pairs}}\\\cline{3-6}
          & & \textsc{Flagged} & \textsc{Exercised}
            & \textsc{WR} & \textsc{BE}\\\hline\hline,
        late after line=\\\hline,
        table foot=]%
        {data/RQ2_p2.csv}{}%
        {\colid & \colnseed & \coladd & \colhit & \coltiem & \coltieb}%
      }%
\end{table}

\paragraph{Deviation basic blocks (\ref{p2}).}
\windranger{} marks a basic block as a DBB when some successor drops reachability to a target. \ourtool{} adds two conditions: loop-gating latches, whose back edge only re-iterates, and
mandatory re-entry chokepoints, which a successor must loop back through to reach the target. Statically \ourtool{} flags 487 blocks across the 10 images, 135 of which \windranger{}'s
criterion does not flag, with up to 26 of them exercised per image. Neither size nor membership is the point. Because the directed distance aggregates over the critical blocks a seed hits,
$D = 1/\sum_i 1/d_i$, every block a classifier declines to mark is a block that cannot separate two seeds: if the blocks that distinguish a seed which approached the target from one that
did not are left unmarked, both receive the same distance and the same energy, and the search has no reason to prefer the one that got closer. Table~\ref{tab:rq2_p2} measures that directly. Over the same queues we recompute the same structural distance twice, once over \ourtool{}'s DBBs and once over only those \windranger{}'s criterion also flags, with the satisfiability adjustment of \ref{p3} disabled in both arms so that the DBB set is the only variable. We then pair every seed that reached the target basic block with every seed that did not, and call a pair tied when the distance gives both the same value, since the two then draw the same energy and the schedule can never prefer the one that reached. Across the subset the median tied
share falls from $41\%$ to $6.6\%$, and no image is worse, three of them unchanged because the extra conditions flag no new blocks there. The extreme is image~15, where without the added blocks the distance is constant over the whole queue and ties every pair, against $8.5\%$ with them.

\begin{table*}
    [!tp]\centering\small
    \setlength{\tabcolsep}{5pt}
    \renewcommand{\arraystretch}{1.2}
    \caption{Left, the distinct values the structural and the satisfiability-adjusted distance each take over the same queue, and the share of tied seed pairs the adjustment splits. Right, the inputs queued on satisfiability alone: their share of the queue, then the share of seeds descending from one, first among seeds that reached the target, then among all queued seeds.}
    \label{tab:rq2_p3}
    \resizebox{\textwidth}{!}{%
        \csvreader[separator=semicolon, head to column names, head to column names prefix=col,
          tabular=|l|r|cccc|ccc|,
            table head=\hline
              \textsc{ID} & \textsc{Seeds} & \multicolumn{4}{c|}{\textsc{Distance resolution}}
                & \multicolumn{3}{c|}{\textsc{Satisfiability-only saves}}\\\cline{3-6}\cline{7-9}
              & & \textsc{Structural} & \textsc{Adjusted} & \textsc{Ratio} & \textsc{Ties split}
                & \textsc{Share of queue} & \textsc{In target lineage} & \textsc{In any lineage}\\\hline\hline,
          late after line=\\\hline,
          table foot=]%
          {data/RQ2_p3.csv}{}%
          {\colid & \colnseed & \coldstruct & \coldadj & \colgain & \colsplit & \colsatq & \colsattgt & \colsatbase}%
        }%
\end{table*}

\paragraph{Satisfiability-adjusted distance (\ref{p3}).}
Structural distance is a function of \emph{which} blocks a seed reached, so two seeds covering the same blocks are tied by construction, however near one came to passing the guard that separates it from the target. The satisfiability discount, $d_i^\star = d_i/(1 + 99\sigma_i)$, makes the value depend on the seed's own comparison operands, turning a path-determined
metric into an input-determined one. Table~\ref{tab:rq2_p3} reports what that buys over the same queues. Its left half counts how many distinct values each metric takes, their ratio, and the share of seed pairs tied under the structural distance that the adjustment separates; its right half isolates the inputs \ourtool{} keeps for advancing satisfiability alone, with no new coverage, giving their share of the queue and how often seeds descend from one, first among the seeds that reached the target and then among all queued seeds. Distinct values are the blunt measure, since seeds sharing a value draw the same energy: on ID~15 the structural distance takes six values across 521 queued seeds while the adjusted one takes 143, a ratio of 23.8. On eight of the 10 images the adjustment separates more than half of the seed pairs the structural distance ties. On the three images where the classifier adds no block (IDs~8, 12, 27) the adjustment splits ties. Resolution is not the whole of it: the satisfiability-only inputs, which exist only because of the adjustment, are a median 5.3\% of the queue, yet on six of the 10 images the seeds that reached the target descend from one more often than the queue as a whole does: 43.2\% of target-reaching seeds against 13.2\% of all queued seeds on ID~8, 53.6\% against 17.6\% on ID~1, and 60.3\% against 49.1\% on ID~31; on ID~4 every target-reaching seed descends from one.
  
Both mechanisms are examined up close in the case studies of Section~\ref{sec:eval:rq5}, where a single target shows which seed the finer distance promotes, and what it buys.

\begin{answer}{2}
    \ourtool{} triggers all 15 targets of the instrumented subset, against 12 for \textsc{Firm-AFL-WR} and 10 for \textsc{Firm-AFL-Go}. Its refined DBBs cut the share of inputs the distance cannot tell apart from $41\%$ to $6.6\%$, and the satisfiability adjustment splits more than half of what remains. Together they give the scheduler a usable gradient on inputs the structural distance scores identically.
\end{answer}

\subsection{RQ3: Energy Allocation}
\label{sec:eval:rq3}

RQ3 asks whether \ourtool{}'s power scheduling (\ref{p4}) allocates energy better than a single global annealing clock, temporally across deviation basic blocks discovered late in a campaign and spatially across co-located targets. We answer it from the per-visit scheduling state \ourtool{} records in its own runs on the instrumented subset, together with the \textsf{no\_energy} ablation, which disables all four levers at once and falls back to the distance-only power of \aflgo{} and \windranger{}. Since the ablation removes them together, three runs per arm cannot separate their outcome effects, so we report engagement and magnitude per lever and the outcome for the scheduler as a whole.

\paragraph{Metric.}
Table~\ref{tab:rq3_sched} reports, per image, what each of the four levers of Section~\ref{sec:scheduling} did, measuring the division of budget directly rather than inferring it from outcomes. The first two columns cover per-block annealing: the share of DBBs first reached only after the global clock had cooled, and the median temperature their own clock held over their first $t_x$. The next three give the share of scheduling decisions on which saturation decay, target fairness and stall reheating engaged. The last column divides the executions \textsf{no\_energy} needs to reach each target by those \ourtool{} needs with the scheduler enabled, taking the median over runs and the geometric mean over an image's targets, so a
value above $1.00\times$ favours the scheduler. It counts executions rather than wall-clock time, since a power schedule changes how many executions a seed is granted, not how fast an execution runs, and wall-clock measurements dilute that with emulation-speed noise varying by an order of magnitude across these images.

\begin{table*}[!tp]\centering\small
        \setlength{\tabcolsep}{8pt}
        \renewcommand{\arraystretch}{1.3}
        \caption{Per image: share of DBBs first reached after the global clock cooled, with the median temperature their own clock holds, the share of scheduling decisions each override changes, and the ratio of executions to reach the target, scheduler disabled over enabled.}
        \label{tab:rq3_sched}
        \resizebox{\textwidth}{!}{%
        \csvreader[separator=semicolon, head to column names, head to column names prefix=col,
          tabular=|l|cc|ccc|c|,
          table head=\hline
            \textsc{ID} & \multicolumn{2}{c|}{\textsc{DBBs found late}}
              & \multicolumn{3}{c|}{\textsc{Scheduling decisions}}
              & \textsc{Execs to target}\\\cline{2-3}\cline{4-6}
              & \textsc{Share} & \textsc{Their temperature}
                & \textsc{Saturation cap} & \textsc{Target fairness} & \textsc{Stall reheat}
                & \textsc{ratio}\\\hline\hline,
          late after line=\\\hline,
          table foot=]%
          {data/RQ3_sched.csv}{}%
          {\colid & \collate & \coltowncl & \colsat & \colfair & \colreheat & \colratio}%
        }%
    \end{table*}

\paragraph{The global clock is cold for most of the campaign.}
With the cooling window at its default of 10 minutes, the global temperature is below 0.05 after minute 10 and effectively zero within the hour, so for 590 of a campaign's 600 minutes a single clock has stopped discriminating at all. That is not a corner case on this subset: on seven of the 10 images deviation basic blocks are still being discovered after the clock has cooled, 40\% to 58\% of them on six of those seven. Those blocks hold a median temperature between 0.22 and 0.60 inside their own window, against at most 0.002 on the global clock that governs every other seed at that moment. This is precisely the exploration budget a single global clock withholds from a block discovered after the campaign has cooled. On IDs~1, 8, and 12 the temperature column is empty because every deviation basic block is found in the first minutes: the lever has no lateness to correct, and measures as doing nothing.

\paragraph{Where budget is contested, the scheduler wins.}
Eight of the 15 targets are co-located, sharing an image with at least one other target, and it is there that energy has to be divided at all. On the three images that carry them, IDs~4,
25 and 26, the full scheduler reaches their targets in $1.11$ to $1.81\times$ fewer executions than \textsf{no\_energy} does, and up to $2.81\times$ on an individual target. The engagement columns confirm that this is the mechanism and not an incidental difference: the target-fairness override, which favours seeds reaching an under-reached target relative to the most-reached one, engages on $14\%$ of scheduling decisions on image~26, $5\%$ on image~25 and $2\%$ on image~4, and on exactly $0\%$ everywhere else, as it must, since with a single target no target can be the under-served
one. Saturation decay, which lowers the energy ceiling of seeds bound to a target that is already well-reached or has stopped yielding new inputs, is the second half of the same story, capping the budget on $96\%$ and $97\%$ of decisions on the two three-target images, higher than on any other image in the subset. The three co-located targets of CVE-2021-34828 to CVE-2021-34830 are the clearest case: the one the scheduler helps most is reached in 4,200 executions against 11,800 without it, and the image as a whole in $1.43\times$ fewer, which is the starvation the spatial machinery exists to prevent.

\paragraph{Scope of the effect.}
On the seven targets that are alone in their image the two arms are close, at a median of $0.93\times$, and the aggregate over all 15 is correspondingly modest. A redistribution mechanism can only pay where there is something to divide; on a single-target campaign the override never fires, the scheduler reduces to its annealing component, and no redistribution is available to measure. Averaging the two regimes into one ladder mixes campaigns where the mechanism is active with campaigns where it is inert by construction, which is why we report them separately. Four single-target images do cost the scheduler executions, IDs~1, 8, 27 and~31. All have a single target, so the fairness override never fires there and only the temporal levers act; at three runs per arm we read the shortfall as run-to-run variance.

\begin{answer}{3}
  \ourtool{}'s power schedule pays where budget is contested: on images holding several targets it reaches them in $1.11$ to $1.81\times$ fewer executions than a single global annealing clock, and the override that redirects budget away from an already-reached target fires only there. Up to $58\%$ of deviation blocks are found only after that global clock has cooled. Where a campaign has one target there is nothing to redistribute and the two are comparable.
\end{answer}

\subsection{RQ4: Component Contribution}
\label{sec:eval:rq4}

RQ4 quantifies how much each of \ourtool{}'s new components contributes to its overall performance. Whereas RQ2 and RQ3 examine two components in depth, RQ4 takes the complementary breadth view: a leave-one-out ablation study that disables each new component in turn and measures what its removal costs in executions to trigger a target.

\begin{table*}[!tp]
     \centering\small
      \setlength{\tabcolsep}{4pt}
      \renewcommand{\arraystretch}{1.2}
      \caption{Per arm: targets triggered; best, geometric mean and worst ratio of executions with the component removed over the full arm; effect size, and the count of targets on which keeping the component wins against those on which removing it wins.}
      \label{tab:rq4_results}
      \resizebox{\textwidth}{!}{%
        \csvreader[separator=semicolon, head to column names, head to column names prefix=col,
          tabular=|l|cc|ccc|cc|,
          table head=\hline
              \textsc{Arm} & \multicolumn{2}{c|}{\textsc{Targets}}
                & \multicolumn{3}{c|}{\textsc{Execs to target ratio}}
                & $\hat{A}_{12}$ & \textsc{Win/loss}\\\cline{2-3}\cline{4-6}
              & \textsc{Triggered} & \textsc{In all runs}
                & \textsc{Best target} & \textsc{Geometric mean} & \textsc{Worst target}
                & & \\\hline\hline,
          late after line=\\\hline,
          table foot=]%
          {data/RQ4_arms.csv}{}%
          {\colarm & \coltrig & \colallthree & \colrmin & \colfac & \colrmax & \colaeff & \colwin}%
        }%
  \end{table*}

\paragraph{Ablation setup.}
The seven arms, six disabling exactly one component and one running the full system, are named in the first column of Table~\ref{tab:rq4_results}; their collection (three runs of 10~h
each) is described in Section~\ref{sec:eval:setup}. The \textsc{targets} columns count how many of the 15 an arm triggers at all and in all three runs. The three middle columns are ratios of executions, the arm's per-target median divided by the full configuration's, so that $1.0$ is parity and a value above it means the removal cost executions: the \textsc{geometric mean} over all 15 targets, with \textsc{best} and \textsc{worst target} the extremes of that same ratio, the target the removal hurt least and the one it hurt most. As in Section~\ref{sec:eval:rq3} we count executions rather than wall-clock time, since every ablated component is scheduling or mutation logic whose own cost is negligible beside an emulated
execution, while emulation throughput varies by an order of magnitude between runs of the same configuration. $\hat{A}_{12}$ is the median per-target effect size, oriented so that values above $0.5$ mean the removed component was pulling weight, and \textsc{Win/loss} counts the targets on which the effect size favours the full arm against those on which it favours the ablated one. Every ratio is formed target by target before it is aggregated, so each comparison holds the target fixed; taking one arm's median target against another's would compare two different targets. Those two extremes show how concentrated the effect is: a component can be decisive on one target and idle on another, which the geometric mean averages away.

\paragraph{Reading the ladder.}
Every arm reaches all 15 target basic blocks, so reachability separates nothing here and the table reports what each removal costs instead. One mechanism stands clear of the rest: disabling satisfiability-aware queue ordering costs a factor of $2.92$ in executions, with the full arm favoured on 14 of the 15 targets against one, costs $22.8\times$ on the target that depends on it most, and costs three targets their all-run reliability. This is consistent with the design: ordering acts on
\emph{which} seed is fuzzed next, and every other mechanism operates on the seed that choice produces, so it is the one removal the rest cannot route around.

\paragraph{Marginal cost is not the same as no contribution.}
The remaining five arms barely move the aggregate, and reading the table by the geometric mean alone would rank them as inert. The per-target columns say otherwise. Removing length extension costs a factor of $1.19$, yet the full configuration is faster on 10 of the 15 targets against five, and on the worst of them the removal costs $28.7\times$; removing the Satisfiability-Preserving Mask (SPM) protection costs $1.20$, with the full arm favoured on 11 targets against four, and costs $14.7\times$ at its worst. Both are the expected signature of a component with redundant substitutes, where removing it lets another mechanism cover the same ground more slowly: many small gains, one target that depends on it outright, and no shift in the middle of the distribution, which is exactly what a single aggregate scalar discards. The satisfiability-adjusted distance is the clearest case of this, at a factor of $1.00$ and yet favoured on nine targets against six, with a worst target of $10.6\times$. The energy arm reproduces the result of Section~\ref{sec:eval:rq3}, where the scheduler's benefit is concentrated on co-located targets and is invisible in an aggregate over campaigns that mostly have nothing to redistribute; it is also the only arm whose worst target costs under $3\times$.

\begin{answer}{4}
    Satisfiability-aware queue ordering is the component that matters most: removing it costs $2.92\times$ more executions, and keeping it is favoured on 14 of the 15 targets. The other components shift the aggregate little, between $0.73$ and $1.20\times$, yet each is decisive on at least one target, where its removal costs up to $28.7\times$.
\end{answer}

\subsection{RQ5: Case Studies}
\label{sec:eval:rq5}

The aggregate results of RQ2 to RQ4 report how often each mechanism helps, not what it does. This section takes three targets from the instrumented subset and follows one mechanism through each:
the DBB classifier (\ref{p2}) on ID~31, the satisfiability-adjusted distance (\ref{p3}) on ID~8, and the power schedule (\ref{p4}) on ID~25. Each target is a distinct sink class, and in each case the decompiled code around the target explains why the aggregate moved.

\paragraph{A deviation block the structural criterion cannot see (ID~31).}
The heap overflow of CVE-2019-11418 sits inside the request loop of \texttt{httpd\_main} (Figure~\ref{lst:cs1}). Both blocks that dominate the target are header tests whose failing branch is a \texttt{continue}: control returns to the loop head, which reaches the target on a later
iteration. Neither successor loses reachability, so \windranger{}'s criterion classifies neither as
a DBB. \ourtool{}'s re-entry condition does: with the block removed, the loop head can no longer reach the target at all, while the fall-through still can. Whether that distinction matters is visible in the corpus: on a replay of the 391 seeds the three campaigns queued on this image, the guarding \texttt{strstr} block is hit by all 121 that reach the target and by 15.2\% of the rest. The 22 blocks \windranger{} flags here fire on over 91\% of the seeds, whether or not they reach the target, so none of them tells the two groups apart. Every \windranger{} block is among the 39 that \ourtool{} flags, and restricted to \windranger{}'s subset the metric leaves 8.3\% of seed pairs tied where the full set leaves none.

\begin{figure}[pos=htbp]
  \begin{lstlisting}[language=C]
    while (wfgets(line, ...)) { /* loop head */
      if (strncasecmp(line, "SOAPAction:", 11))
        continue; /* DBB */
      v = line + 11;
      v += strspn(v, " \t");
      if (!strstr(v, "HNAP1"))
        continue; /* DBB */
      p = malloc(1024); /* target */
      memset(p, 0, 1024);
      memcpy(p, v, strlen(v));
    }
    \end{lstlisting}
    \caption{Header parse loop of \texttt{httpd\_main} (ID~31).}
    \label{lst:cs1}
\end{figure}

\paragraph{Distance resolution when every seed hits the same blocks (ID~8).}
CVE-2025-6328 is reached through a four-state cookie parser driven by an environment variable (Figure~\ref{lst:cs2}). The state machine is a single loop that every well-formed seed enters, so the blocks an execution covers barely depend on the input: replaying the $492$ seeds the campaigns queued on this image, eight of the nine blocks they exercise are hit by $94\%$ to $100\%$ of them. Block coverage alone therefore cannot rank this queue, and the structural aggregate takes three distinct values across the corpus. CmpLog cannot supply the answer either: the comparison receives a string object, so the operand logged against \texttt{uid} is its header, not the name. What separates a seed here is not which blocks it covered but how far it advanced through the cookie name, which is what the satisfiability term measures. On the same seeds the adjusted distance takes $41$ distinct values and splits $61.7\%$ of the pairs the structural metric leaves tied. Inputs saved only because their satisfiability improved are $3.0\%$ of the corpus, yet ancestors of $43.2\%$ of the seeds that reach the target against $13.2\%$ of the corpus.

\begin{figure}[pos=htbp]
    \begin{lstlisting}[language=C]
    cookie = getenv("HTTP_COOKIE");
    for (p = cookie; *p; p++) { /* loop head */
      switch (state) {
      case 1:
        if (*p == '=') { state = 2; continue; }
        sobj_add_char(name, *p);
        continue;
      case 2:
        if (*p != ';') {
          sobj_add_char(val, *p);
          continue;
        }
        state = 3;
      case 3:
        s = sobj_get_string(val);
        if (sobj_strcmp(name, "uid") == 0)
          sobj_add_string(sess, s); /* target */
      }
    }
    \end{lstlisting}
    \caption{Cookie state machine of \texttt{sess\_get\_uid} (ID~8).}
    \label{lst:cs2}
\end{figure}

\paragraph{Dividing energy across three co-located sinks (ID~25).}
The three targets of ID~25 are three unchecked copies in one \texttt{http\_request\_parse} dispatch chain (Figure~\ref{lst:cs3}), and they are not equally hard. The two header sinks are reached as soon as the header name is right, while the cookie sink needs both the header and the literal \texttt{uid=} inside its value, which costs two orders of magnitude more executions. This is the regime a single global annealing clock handles worst: it cools within the first 10 minutes, after which the two cheap targets are already saturated and keep absorbing budget that the expensive one never receives. On this image the target-fairness override engages on $5\%$ of scheduling decisions and saturation decay caps the budget on $97\%$ of them, both of which are inert in \aflgo{} and \windranger{} by construction. Turning the scheduler off costs $2.2\times$, $1.1\times$ and $2.5\times$ the median executions on the three targets, and the cookie sink is then reached in only
two of the three runs.

\begin{figure}[pos=htbp]
    \begin{lstlisting}[language=C]
    if (!strcasecmp(hdr, "HNAP_AUTH")) {
      memset(g_auth, 0, 64);
      strcpy(g_auth, val); /* target */
    } else if (!strcasecmp(hdr, "SOAPAction")) {
      memset(buf, 0, 256);
      strcpy(buf, val); /* target */
    } else if (!strcasecmp(hdr, "Cookie")) {
      strcpy(ck, val);
      if ((m = strstr(ck, "uid=")))
        strcpy(ck, m + 4); /* target */
    }
    \end{lstlisting}
    \caption{Dispatch chain of \texttt{http\_request\_parse} (ID~25).}
    \label{lst:cs3}
\end{figure}

\begin{answer}{5}
The case studies show each mechanism acting on a real target: a deviation block the structural criterion cannot see (\ref{p2}), a distance separating seeds that cover identical blocks (\ref{p3}), and energy divided across three co-located sinks (\ref{p4}). Each is the behaviour RQ2--RQ4 report in aggregate, observed on one target.
\end{answer}

\section{Related Work}
\label{sec:related}

\paragraph{Directed greybox fuzzing.}
Beyond \aflgo{} and \windranger{} (Section~\ref{sec:back:dgf}), subsequent work has sought to sharpen the distance metric and to reduce the cost of instrumenting code that is irrelevant to the target. \textsc{Hawkeye}~\cite{Hawk} sharpens \aflgo{}'s aggregation with a function-level similarity term and an adaptive mutation policy, but still allocates energy from one seed-level score pooled over all targets (\ref{p4}). \textsc{DAFL}~\cite{dafl} observes that structural proximity blends blocks that merely lie near the target with those that actually influence whether it is reached; it derives distance from \emph{data-dependency} relations, selectively instrumenting only the functions that the target's definition-use chains reach, thereby suppressing the noise that arithmetic-mean control-flow distance introduces. This semantics-aware notion of relevance is close in spirit to \ourtool{}'s satisfiability distance, yet \textsc{DAFL} operates on source-level data-flow analysis and, like its predecessors, remains inapplicable to the closed-source binaries that dominate firmware. \textsc{LibAFLGo}~\cite{geretto2025libaflgo} takes a complementary engineering approach by implementing directed fuzzing over the modern \textsc{LibAFL} framework. Its aim is to demonstrate how far the DGF literature has drifted from the coverage-guided fuzzing community's recent advances by continuing to build on \aflgo{}'s ageing codebase, a gap the authors argue has hindered the real-world adoption of DGF; they report that on such a stack no directed policy outperforms undirected \textsc{LibAFL}. \ourtool{} shares this concern but acts on it along a different axis: rather than porting the search to a new framework, it first strengthens the input-generation stack of its baseline fuzzer, by porting CmpLog operand resolution and grammar-aware mutations (Section~\ref{sec:mut-opt}), before layering directed scheduling on top, and it does so for a domain \textsc{LibAFLGo} does not address: emulated, closed-source firmware. 

Other efforts refine reachability and target relevance along independent axes: \textsc{SelectFuzz}~\cite{luo2023selectfuzz} instruments only the path-relevant blocks, defined as those on which the target is control or data dependent, so that coverage feedback is not diluted by blocks that cannot influence whether the target is reached; \textsc{Beacon}~\cite{beacon} derives lightweight preconditions by backward interval analysis and injects them as assertions that abort a run early once its path is provably unable to reach the target; \textsc{FishFuzz}~\cite{zheng2023fishfuzz} scales direction to thousands of targets by favouring the closest seed  to each \emph{not yet exhausted} sink rather than pooling every target into one score; \textsc{TransferFuzz}~\cite{transferfuzz} replays historical execution traces of a known vulnerability to steer testing toward the same code once it has been reused or propagated into downstream software, and also abandons \aflgo{}'s single clock, annealing per function-level state along a trace recorded from the binary the CVE names; \ourtool{} anneals per DBB, with no prior binary required (\ref{p4}); and \textsc{TriFuzz}~\cite{trifuzz} couples a loop-aware probabilistic distance metric with symbolic execution applied selectively at the hard branches the distance heuristic alone cannot solve, absorbing loop structure into the metric where \ourtool{} instead corrects the DBB classifier (\ref{p2}). All of these, however, assume source availability or compile-time instrumentation and are evaluated on conventional software rather than firmware.

\paragraph{Firmware fuzzing.}
Directed fuzzing has only recently begun to reach the firmware domain surveyed in Section~\ref{sec:back:firmware-fuzzing}, with the two closest efforts differing from \ourtool{} in
\emph{what} they steer toward and \emph{how precisely}. \textsc{ReachDFuzz}~\cite{reachdfuzz} brings a directed strategy to IoT firmware by using static analysis to flag dangerous library calls (for instance, a \texttt{strcpy}) that external input can reach, then discarding inputs whose execution path can never arrive at one of them, reporting fewer wasted executions and better bug detection than \firmafl{}~\cite{firmafl}. Its notion of \emph{close to the target} is binary: a block either can reach a target block or cannot, so every input that survives pruning looks equally good to the fuzzer. \ourtool{}, by contrast, grades proximity by shortest-path distance, so it can tell two surviving inputs apart when only one is edging closer to the vulnerable statement, giving the search a gradient to follow. A second complementary line of work asks a different question: rather than steering toward a target the analyst has \emph{already} chosen, \citet{qin2025dgf} focus on \emph{finding} the targets in the first place. They fuzz power terminal firmware and use binary-code-similarity comparison, matching the firmware's code against known-vulnerable code patterns, to locate candidate vulnerable sites automatically, even across different CPU architectures, without a human pointing them out. Their own scheduling then targets blocks at \emph{function} granularity, so it stays complementary to, and could feed directly into, the basic-block distance \ourtool{} contributes once
a target is known. \textsc{Labrador}~\cite{liu2024labrador} steers devices that resist emulation entirely, inferring both coverage and distance to a sink from network responses rather than from instrumentation; its feedback is response-level, and applies exactly where \ourtool{}'s emulation-based pipeline cannot run. Closest of all, \textsc{FirmAgent}~\cite{firmagent} scores firmware blocks by shortest-path proximity to pattern-matched sink APIs, but to discover taint sources for an LLM agent rather than to reach an analyst-chosen target, reporting neither a comparison against directed baselines nor a time-to-exposure. To our knowledge, \ourtool{} is the first framework to schedule closed-source firmware fuzzing by basic-block-level distance to user-specified targets.

A parallel line of work tackles a different axis, focusing on the execution engine rather than direction: \textsc{Greenhouse}~\cite{jun2023greenhouse} automatically re-hosts firmware for high-throughput fuzzing, broadening the emulation coverage of the hybrid schemes on which \ourtool{} builds, but it remains coverage-guided and offers no notion of a target. \textsc{HouseFuzz}~\cite{housefuzz} and \textsc{EquAFL}~\cite{equafl} likewise advance service-aware and emulation-efficient firmware fuzzing without a directed component.

\paragraph{Large language models for fuzzing.}
A recent line of work applies large language models (LLMs) to synthesise the structured inputs that random mutation struggles to produce. \textsc{ChatAFL}~\cite{llm3} interrogates an LLM for the machine-readable message grammar of a protocol, and uses it to construct per-message-type grammars, mutate messages, and predict the next message in a stateful sequence, substantially improving state and code coverage over unguided protocol fuzzers. \textsc{ChatHTTPFuzz}~\cite{yang2025chathttpfuzz} applies the same idea to the IoT setting most relevant to \ourtool{}, using an LLM to interpret firmware HTTP interfaces and generate well-formed requests that survive early input validation. Both, however, employ the model to broaden coverage rather than to steer execution toward a specific target. Other work uses LLMs for input generation more generally: \textsc{LLMIF}~\cite{llmif} augments a model with protocol specifications to fuzz IoT devices, and \textsc{Fuzz4All}~\cite{fuzz4all} generates inputs across many languages by autoprompting a single model. Most relevant to  \ourtool{}, \textsc{ISC4DGF}~\cite{isc4dgf} applies an LLM to \emph{directed} fuzzing, generating an optimised initial seed corpus that reaches target sites faster on source-available benchmarks. \ourtool{} uses an LLM for the same purpose, seeding the campaign, but on closed-source firmware, and it additionally distills the LLM's knowledge of the input format into a per-binary grammar that then drives structure-aware mutation throughout the run, rather than confining the model's contribution to the initial corpus.

\section{Limitations}
\label{sec:limitations}

\paragraph{Static binary analysis.}
\ourtool{}'s distance metric is only as complete as the control-flow and call graphs recovered from the stripped binary. It relies on a commercial disassembler~\cite{ida} to reconstruct these graphs, and binary-level graph recovery is inherently imprecise: disassembly of optimised, stripped, or obfuscated code frequently misidentifies instruction and function boundaries~\cite{andriesse2016disasm}, an imprecision \ourtool{} inherits, as do other binary-level directed fuzzers~\cite{uaf}. The most consequential gap is the resolution of indirect transfers: calls through function pointers and computed jumps. The targets of such transfers cannot in general be recovered statically~\cite{typearmor,svf}. As described in Section~\ref{sec:binary-pipeline}, \ourtool{} attempts to resolve these dynamically by adding edges when a transfer is observed at runtime. A target reachable exclusively through an indirect edge that the fuzzer never exercises therefore remains at infinite distance and is never prioritised by the scheduler. Finally, selecting the target basic blocks is itself a manual step: for each new campaign an analyst must read the disassembly and pinpoint the exact block corresponding to the vulnerable sink, a process that is both time-consuming and error-prone, as a misread offset silently directs the fuzzer to the wrong location; automated target discovery~\cite{qin2025dgf} is complementary and could supply this input. This manual static-analysis effort, rather than any property of the search itself, is what bounds the scale at which new campaigns can be set up.

\paragraph{Emulation fidelity and coverage.}
\ourtool{} is built on \firmafl{}'s augmented process emulation~\cite{firmafl} and consequently inherits the limitations of the Firmadyne rehosting pipeline~\cite{firmadyne} on which it depends. Firmadyne successfully boots only a fraction of real-world firmware images~\cite{kim2020firmae}, so the set of targets \ourtool{} can analyse is bounded by what the emulator can bring up rather than by the fuzzer itself. Furthermore, \firmafl{}'s process-emulation throughput is no longer state of the art: more recent rehosting systems achieve broader device support and higher fidelity~\cite{jun2023greenhouse,equafl,housefuzz}, and one of them executes faster than our back-end by a wide margin (Section~\ref{sec:eval:rq1}). Three further constraints follow from the execution model. First, \ourtool{} injects a single test case at one fork point, on whichever of its five input channels the target uses, so vulnerabilities that require a multi-message session, such as an authentication handshake followed by a stateful request sequence, are not modelled, as they are by stateful protocol fuzzers~\cite{aflnet,llm3}. Second, \ourtool{} targets Linux-based MIPS only, though \firmafl{} also supports ARM, and not bare-metal or RTOS firmware, for which dedicated peripheral-modelling approaches exist~\cite{feng2020p2im,scharnowski2022fuzzware,hernandez2022firmwire}. Third, for a minority of targets the execution time is dominated by QEMU's dynamic binary translation~\cite{qemu} rather than by system-call handling, which caps the throughput gain that directed scheduling can deliver.

\paragraph{Experimental validity.}
Each (configuration, target) pair is repeated 10 times, matching the practice of recent directed fuzzers~\cite{beacon}, but falling short of the more than 20 repetitions recommended by a dedicated study of directed-fuzzing evaluation~\cite{kim2024evaluating}, which shows that time-to-exposure estimates stay sensitive to randomness below that threshold; we instead used the budget for breadth, 40 targets over 32 images, rather than further repetitions, with RQ1 alone costing 20{,}000 CPU-hours across the five \firmafl{}-based configurations.

\paragraph{Target reachability.}
Directed fuzzing presupposes that a feasible input path to the target exists, yet many firmware vulnerability sites are guarded by conditions that fuzz-controlled input cannot satisfy, including authentication checks, session tokens, or LAN subnet restrictions on the requesting client.  Handling such gates is an open problem acknowledged by contemporary firmware fuzzers~\cite{housefuzz}, and path-pruning directed fuzzers~\cite{beacon,luo2023selectfuzz} are equally unable to reach a target whose guard is not input-satisfiable.  \ourtool{}'s comparison-guided mutation~\cite{cmplog,redqueen} resolves many such guards by matching logged operands, but it cannot bootstrap a match when the operand bytes fall outside the colorized taint region, nor when the guard is a checksum or an encoded transform of the input. When multiple targets are specified, the power schedule must divide energy among them, and closer or more easily reached targets can starve the remainder; \ourtool{}'s per-target scheduling mitigates but does not eliminate this effect, a difficulty also observed in many-target directed fuzzing~\cite{zheng2023fishfuzz}.

\paragraph{Grammar and LLM-assisted generation.}
\ourtool{} uses a locally hosted large language model~\cite{qwen25,llm2} to derive per-binary input grammars and seeds, which introduces the well-documented reliability concerns of LLM-based tooling: outputs are non-deterministic and may be hallucinated, requiring validation before use~\cite{llm1}. For soundness, the generated grammars are deliberately generic and structural rather than target-specific, and they cover only the environment variable and file stream input channels; other inputs fall back on general-purpose grammar mutation~\cite{grammar_mutators,wang2019superion,nautilus} via the Grammar Mutator component, without model assistance.

\paragraph{Bug oracle.}
\ourtool{} detects memory-safety violations (heap and global-variable overflows) by hooking the libc copy routines and checking each write against a \texttt{.dynsym} object bound, an allocation redzone, or the uClibc~\cite{uclibc} allocator's chunk metadata (Section~\ref{sec:bug-oracles}), because closed-source firmware binaries cannot readily be recompiled with the sanitisers~\cite{asan,song2019sok} that greybox fuzzers ordinarily rely on for bug detection. Detection is therefore bounded by what the binary declares: stack buffers and objects without a \texttt{.dynsym} entry have no recoverable size and fall back to a length threshold, so smaller overflows of those go unreported, as do use-after-free, double-free, and \texttt{mmap}-backed allocations. Consequently, vulnerability classes that do not manifest as memory corruption, such as command injection, information disclosure, authentication bypass,
and logic errors, are not surfaced as crashes and fall outside \ourtool{}'s current detection scope.

\section{Conclusions}
\label{sec:conclusions}

Firmware images routinely reuse third-party components with known vulnerabilities, and checking whether a given device actually exposes one of them means steering the search toward the suspected code rather than exploring the binary at large. Directed greybox fuzzing does exactly this, but has reached firmware only at function granularity, too coarse to aim at the vulnerable block itself: it assumes source code, which firmware does not provide, and even where it has been ported to binaries, its guidance breaks down on the loops, hard comparisons, and multiple co-located sinks that firmware protocol parsers make routine.

\ourtool{} makes directed fuzzing work under these conditions. It recovers control-flow and call graphs directly from stripped MIPS binaries, refining them at runtime as indirect calls are observed, so no source code or compiler support is needed. It classifies critical branches with a loop-aware criterion that catches guard conditions a purely reachability-based test misses. It replaces control-flow-hop counting with a satisfiability score computed from the comparison operands a seed actually produces, so a seed one byte away from passing a hard check is prioritised over one that is merely closer on the graph. And it schedules energy per branch and per target rather than off a single global clock, so a branch discovered late in the campaign, or a target overshadowed by an easier one, still gets a fair share of exploration. Supporting mechanisms round out the system: grammar- and LLM-assisted mutation keeps generated inputs structurally valid, the satisfiability-preserving mask protects comparisons a seed has already solved from being undone by later mutation, length-extension mutation grows inputs past buffer bounds to trigger overflows, and two oracles catch overflows that would otherwise leave no crash signal.

On 40 vulnerability sites across 32 real IoT firmware images, \ourtool{} reproduces every target within a 10~h budget, more than any baseline we compare against, and does so faster by a geometric mean of $9.5\times$ to $72.5\times$. Instrumented runs confirm this is not an artefact of aggregation: the branch classifier resolves ties a reachability-only test cannot, the satisfiability score separates seeds a purely structural distance cannot tell apart, and the scheduler visibly redirects budget toward starved targets whenever a campaign has more than one.

These results come with real limits, detailed in Section~\ref{sec:limitations}: distance is only as good as static disassembly and runtime call-graph discovery allow, target selection is still a manual step, and bug detection is limited to the memory-safety violations the binary-level oracles can see. Extending \ourtool{} with automated target discovery, faster re-hosting back-ends, and support for stateful multi-message protocols is a natural next step.


  

\balance
\bibliographystyle{cas-model2-names}
\bibliography{references}



\end{document}